\documentclass[aps,prd,english,superscriptaddress,11pt,notitlepage]{revtex4}
	\usepackage[colorlinks=true, a4paper=true, pdfstartview=FitV,
linkcolor=blue, citecolor=blue, urlcolor=blue]{hyperref}

\usepackage{amsmath}
\usepackage{amssymb}
\usepackage{graphicx}
\usepackage{hhline}
\usepackage{textcomp}
\makeatletter
\usepackage{babel}
\newcommand{\bea}{\begin{eqnarray}}
\newcommand{\eea}{\end{eqnarray}}

\newcommand{\be}{\begin{equation}}
\newcommand{\ee}{\end{equation}}
\usepackage[active]{srcltx}
\begin{document}


 \title{Spectral walls in multifield kink dynamics}
 
 \author{C. Adam}
\affiliation{Departamento de F\'isica de Part\'iculas, Universidad de Santiago de Compostela and Instituto Galego de F\'isica de Altas Enerxias (IGFAE) E-15782 Santiago de Compostela, Spain}
\author{K. Oles}
\affiliation{Institute of Physics,  Jagiellonian University, Lojasiewicza 11, Krak\'{o}w, Poland}
\author{T. Romanczukiewicz}
\affiliation{Institute of Physics,  Jagiellonian University, Lojasiewicza 11, Krak\'{o}w, Poland}
\author{A. Wereszczynski}
\affiliation{Institute of Physics,  Jagiellonian University, Lojasiewicza 11, Krak\'{o}w, Poland}
\author{W. J. Zakrzewski}
\affiliation{Department of Mathematical Sciences, University of Durham, Durham DH1 3LE,
United Kingdom}
 
\begin{abstract}
We show that spectral walls are common phenomena in the dynamics of kinks in (1+1) dimensions. They occur in models based on two or more scalar fields with a nonempty Bogomol'nyi-Prasad-Sommerfield (BPS) sector, hosting two zero modes, where they are one of the main factors governing the soliton dynamics.
We also show that spectral walls appear as singularities of the dynamical vibrational moduli space. 
\end{abstract}

\maketitle

\section{Introduction}

Despite their prevalence in many physical systems, there is no satisfactory understanding of interactions of topological solitons in generic, non-integrable models. Even in the $\phi^4$ model in (1+1) dimensions, which is probably the simplest, prototypic theory with topological kinks, there is no quantitative explanation of the involved, chaotic pattern of final state formation in kink-antikink collisions, like soliton annihilation or their backscattering after several bounces, although a significant role of the normal modes has been conjectured \cite{kevrekidis} (see, however, \cite{MORW} for very recent progress on this issue). There are, in fact, three qualitatively distinct ways in which solitons can interact. First of all, depending on their topological charge and relative orientation, solitons attract or repel each other, deforming their shapes. Secondly, most solitons carry internal degrees of freedom (DoF), i.e., normal modes, which can be easily excited during collisions. The transfer of energy from the kinetic to the internal DoF can further complicate the soliton dynamics. Finally, solitons couple to radiation, which also may lead to the appearance of important phenomena (like, e.g., negative radiation pressure). All these factors contribute to the fascinating complexity of solitonic interactions. 

A common strategy which may provide further insight into the complicated dynamics of solitonic collisions is to reduce the full field theory (a system of partial differential equations with infinitely many DoF) to a mechanical like system with a finite number of DoF, which evolve via ordinary differential equations \cite{NM-moduli}. These reduced effective DoF, called {\it moduli}, span a subset of field configurations which should capture the main aspects of the dynamics. Of course, one of the main issues is to identify the proper moduli. A natural and simply guess is to consider the lightest modes, that is, the zero modes (kinetic DoF which, e.g., may encode the information about the inter-soliton distance) and massive normal modes (internal DoF). This prescription works very well if a soliton-soliton collision occurs in the so called Bogomol'nyi-Prasad-Sommerfield (BPS) sector, i.e., where the static intersoliton force vanishes \cite{SM}. Thus, the (static) initial and final states are connected by a path in the space of field configurations where all points (field configurations) have exactly the same energy. While this strategy can be extended to near-BPS scattering \cite{Manton-1}, there is still an ongoing debate on how to identify the correct moduli for a soliton-antisoliton annihilation in a generic non-BPS process. 

However, even for a BPS scattering, where the canonical moduli space exists, the number of effective DoF (and therefore the dimension of the moduli space) can vary during the collision. This is related to the fact that the structure of the normal modes changes while the solitons approach each other. In particular, a normal mode can hit the mass threshold and disappear into the continuum spectrum. Of course, one can ask whether such a change in the number of the effective DoF (normal modes) has any impact on the soliton dynamics. Recently, an affirmative answer to this question was reported in the case of BPS soliton-impurity systems, that is, in the framework of the {\it BPS impurity models} \cite{imp-phi4}-\cite{SD-phi4}, where the inter-soliton force is effectively switched off by a properly chosen external field (impurity). Indeed, the disappearance of a normal mode (strictly speaking its transition to the continuous spectrum at a certain value of a moduli coordinate $b_{sw}$ related to the inter-soliton distance) implies the appearance of the {\it thin spectral wall} (SW) phenomenon \cite{spectral-wall}. A spectral wall is a spatially well defined obstacle which affects the soliton dynamics, where the scattering solitons can form a stationary, in principle arbitrarily long-living, oscillating state. The actual behaviour depends on the corresponding {\em amplitude} of the mode crossing the mass threshold. At a critical value of this amplitude, the solitons are confined at the SW forming the stationary solution. If the mode is less excited, the solitons pass through the wall. For larger excitations they are reflected back where the reflection happens sooner (at larger mutual distance) as the amplitude grows.
Concretely, the SW phenomenon has been investigated both for a kink-on-impurity scattering \cite{spectral-wall} and for kink-antikink scattering in an impurity background \cite{SD-phi4}. Note that a SW can show up even if the solitons participating in the collision are quite far away from each other. 

Here we confirm, for the first time, the existence of SW as an effect of the transition of the mode to the continuum, in a model which does not require any impurity. More precisely, we find SW in a two-scalar-field theory in (1+1) dim which has a BPS solution where both fields take nontrivial values.
Concretely, one field (let us call it $\phi$) is a kink, whereas the second field ($\psi$) describes a one-parameter family of kink-antikink pairs.
In our analysis, we choose a particular model which is distinguished both by its similarity to the model of Ref. \cite{SD-phi4} and by its simplicity, which allows us to perform many calculations analytically.
 We shall argue, however, that our results are not restricted to this particular model, but are generic for many scalar field theories possessing a nontrivial BPS sector. 
 In the simplest case, all that is required, in addition, is that this BPS sector hosts two zero modes,  where the first (trivial) zero mode is related to a simultaneous translation of the solitons, whereas the second (non-trivial) zero mode encodes the mutual distance between the solitons. The second zero mode is responsible for the vanishing static intersoliton force. These results imply that SW seem to be rather generic phenomena which exist in many BPS two-field models, significantly modifying their dynamics. 

\section{Model}
\subsection{Two-field models and BPS sectors}
We want to study the formation of spectral walls in a nontrivial BPS sector of a scalar field theory with (at least) two fields. 
The crucial properties which should be shared by a candidate two-scalar-field theory are the following.
\begin{enumerate}
\item There should exist a non-empty BPS sector which contains nontrivial field configurations (e.g., solitons) in {\it both} fields, $\phi$ and $\psi$. In our concrete example, $\phi$ will describe a kink (or antikink), and $\psi$ will describe a kink-antikink pair.
\item In the simplest case relevant for our present investigation, solutions of this BPS sector should be parametrized by {\it two} real parameters $a,b$ ({\it moduli}), which give rise to a two-dimensional space of energetically equivalent solutions ({\em moduli space}). Physically, the first parameter $a$ corresponds to a trivial translation of the BPS solutions while the second, nontrivial parameter $b$ controls the relative distance between the solitons in the $\phi$ and $\psi$ fields. In our concrete example, $b$ describes the distance between the kink and the kink-antikink pair symmetrically located about the kink position.  
\item The spectrum of small perturbations should contain a normal mode which crosses the mass threshold for a certain value of the second modulus $b$, corresponding to a certain distance between the solitons in the $\phi$ and $\psi$ fields. 
\end{enumerate}
There are many known standard Lorentz-invariant two-field models sharing the first two conditions, see e.g., \cite{m}-\cite{bazeia1}. As one particular example, let us briefly describe the model introduced in \cite{bazeia2}. One method to find the BPS equations, inspired by supersymmetry, consists in finding the superpotential $W(\phi,\psi)$ for a given field theory potential $V(\phi ,\psi)$, related by (here $W_{,\phi} \equiv (\partial /\partial \phi )W$, etc.)
\be
V = \frac{1}{2} \left( W_{,\phi}^2 + W_{,\psi}^2 \right).
\ee
The BPS equations are then 
\be 
\frac{d}{dx} \phi = W_{,\phi} , \quad \frac{d}{dx} \psi = W_{,\psi}
\ee
and the topological energy bound is
\be
E>E_{top} =\vert \int dx (\phi_{,x} W_{,\phi} + \psi_{,x}W_{,\psi})\vert = \vert W(\phi_\infty , \psi_\infty) - W(\phi_{-\infty}, \psi_{-\infty})\vert
\ee
where the limiting values $\phi_{\pm \infty} = \lim_{x\to \pm \infty} \phi(x)$, etc., must take values in the vacuum manifold of the model. In \cite{bazeia2} a family of models depending on various coupling constants was introduced, but for a particular choice the resulting superpotential is
\be
W = \frac{1}{3}\phi^3 - \frac{1}{4}\phi (1-\psi^2),
\ee
and the vacuum manifold consists of the four points $\phi =0, \psi = \pm 1$, and $\phi = \pm \frac{1}{2}, \psi =0$. In particular, for the boundary conditions $\phi_{\pm\infty} = \pm\frac{1}{2}$, $\psi_{\pm\infty} =0$, corresponding to a $\phi$ kink and a $\psi$ kink-antikink pair, the energy bound is nontrivial, $E_{top} = \vert W(1/2,0) - W(-1/2,0)\vert = (1/6)$. The corresponding BPS solutions exist and depend on two moduli $a,b$, where $a$ describes a translation of the solutions, whereas $b$ describes the relative distance between the kink and the kink-antikink pair. Explicit expressions for these solutions and figures showing the corresponding profiles can be found, e.g., in \cite{izq7,izq4}.   

The normal mode structure of the above model has been studied in \cite{bazeia2} for some isolated solutions (fixed values of the moduli $a$ and $b$). The variation of the normal modes and their frequencies with the second modulus $b$, on the other hand, has not been studied either in the model of \cite{bazeia2} or in other multi-field models with a nontrivial BPS sector, to the best of our knowledge.  The crossing of the mass threshold by a normal mode, however, is a rather generic phenomenon in models where it has been investigated. Indeed, it happens quite generically in theories with one scalar field in an impurity background \cite{imp-phi4}, \cite{spectral-wall}. Therefore, we expect that the phenomena obtained in our work will exist in many of these models. 

\subsection{The model}

In the present work, we shall consider a slightly different model. It consists of two real scalar fields $\phi$ and $\psi$ in 1+1 dimensional Minkowski space and has the following properties.
\begin{itemize}
\item
The BPS equation of the field $\phi$ is {\em independent} of the second field and is, therefore, identical to the BPS equation for a kink (or antikink) in a corresponding one-field model. The second BPS equation for $\psi$ can then be solved in the background of this kink field, describing a kink-antikink pair.  The situation is, therefore, quite similar to the impurity models, although $\phi$ is not an impurity but a dynamical field. The impurity model of \cite{SD-phi4} can, in fact, be recovered by sending the mass scale of the $\phi$ field to infinity, see Section VII.
\item
The static energy functional in $\psi$ is already a complete square. This implies that a spatial derivative term $\partial_x \psi$ appears in the Lagrangian, which breaks Lorentz invariance. As a consequence of this choice, only the completion of the square w.r.t. the $\phi$ field requires the addition of a topological term, i.e., a total derivative. The resulting energy bound, therefore, is trivial in $\psi$ and depends only on the topology (i.e., the boundary conditions) of $\phi$. 
\item
As a consequence of our choices, the spectral problem simplifies. The normal modes and their frequencies can, in fact, be determined analytically in two limiting cases, namely for a widely separated kink-antikink pair ($b \to \infty$), and for complete annihilation ($b=0$). The most important observation is that there are altogether nine normal modes for $b\to \infty$, whereas there are only four modes for $b=0$. This implies that during the evolution from a sufficiently large $b$ to $b=0$, five normal modes {\em must} have crossed the continuum threshold, and the spectral wall phenomenon {\em must} occur.

\end{itemize}

We emphasize, again, that the particular model we consider is by no means necessary for the phenomena we want to discuss, which is the effect of the transition of a normal mode into the continuum spectrum, and similar results can most likely be found for many Lorentz-invariant field theories with nontrivial energy bounds in both fields. It turns out, however, that the calculations for our specific model are particularly simple, and many results can, in fact, be derived analytically.

We assume that certain units of length and energy have been chosen, such that both the space-time coordinates $(t,x)$ and the scalar fields are dimensionless. Our Lagrangian then takes the form
\be
\mathcal{L}[\phi,\psi]= \frac{M}{2} \left(\partial_\mu \phi \right)^2 - \frac{M}{2} (\phi^2-1)^2 +\frac{m}{2} \left(\partial_\mu \psi \right)^2 - \frac{m}{2}  \phi^2(\psi^2-1)^2+m \phi (1-\psi^2)\partial_x \psi, \label{model}
\ee
where the two real scalar field $\phi, \psi: \mathbb{R}\times \mathbb{R} \ni (t,x) \rightarrow \phi(x,t), \psi(x,t) \in \mathbb{R}$ interact via a potential
\be
V(\phi, \psi) =  \frac{M}{2} (\phi^2-1)^2 + \frac{m}{2}  \phi^2(\psi^2-1)^2 \label{potential}
\ee
and via a derivative coupling 
\be
m \phi (\psi^2-1)\partial_x \psi.
\ee
Here, $M$ and $m$ are dimensionless constants which control the relative energy contributions of the two fields. In the subsequent analysis they are set to one. The potential (\ref{potential}) has four vacua $(\phi_v, \psi_v)=(\pm 1, \pm 1)$. The derivative coupling violates Lorentz invariance, but similar couplings between a scalar field and a spatial derivative of another scalar field have been used in realistic systems, see for example the proton transfer in hydrogen-bonded systems \cite{pnevmatikos}. Once again, we stress that our findings will apply to many two-scalar-field models in (1+1) dimensions, provided that a non-empty BPS sector with two zero modes exists.  

The static energy can be bounded from below by a sort of Bogomol'nyi decomposition. Namely:
\bea
E&=&\int_{-\infty}^\infty \left[ \frac{1}{2} \left(\frac{d\phi}{dx} \right)^2 +\frac{1}{2} \left(\frac{d\psi}{dx} \right)^2 + \frac{1}{2} (\phi^2-1)^2 + \frac{1}{2}  \phi^2(\psi^2-1)^2- \phi (1-\psi^2)\frac{d\psi}{dx}  \right] dx \nonumber\\
&=&\int_{-\infty}^\infty \left[ \frac{1}{2} \left(\frac{d\phi}{dx} \pm (1-\phi^2) \right)^2  + \frac{1}{2} \left(\frac{d\psi}{dx} - \phi(1-\psi^2) \right)^2    \mp (1-\phi^2)\frac{d\phi}{dx}  \right] dx \nonumber  \\
&\geq & \left| \int_{-\infty}^\infty  (1-\phi^2)\frac{d\phi}{dx} dx \right| = \int_{-1}^1 (1-\phi^2)d\phi |Q_\phi| = \frac{4}{3} |Q_\phi|, 
\eea
where $Q_\phi$ is the topological charge of the field $\phi$
\be
Q_\phi=\frac{1}{2} (\phi(x=\infty) - \phi(x=-\infty)).
\ee
As we have mentioned before, the topological energy bound does not depend on the topological charge of the second scalar. 
Furthermore, the bound is saturated, if and only if, the following Bogomol'nyi equations are satisfied
\bea
\frac{d\phi}{dx} \pm (1-\phi^2) &=& 0, \\
\frac{d\psi}{dx} - \phi(1-\psi^2) &=& 0.
\eea
The solutions of these equations are called Bogomol'nyi-Prasad-Sommerfield (BPS) solitons. One can easily verify that they satisfy the second order Euler-Lagrange equations. Observe also that there is only one Bogomol'nyi equation for the $\psi$ field (only one sign). Therefore, in contrast to the $\phi$ field, $\psi$ is a kind of chiral scalar. 
We will see below that, as a consequence of this chirality, only some combinations of BPS solutions will show up. We would also like to add that this set of first order equations is identical to the set of equations emerging from the so-called {\it iterated $\phi^4$ model} with $n=2$ \cite{Nick}.  
\begin{figure}
\includegraphics[width=0.6\textwidth]{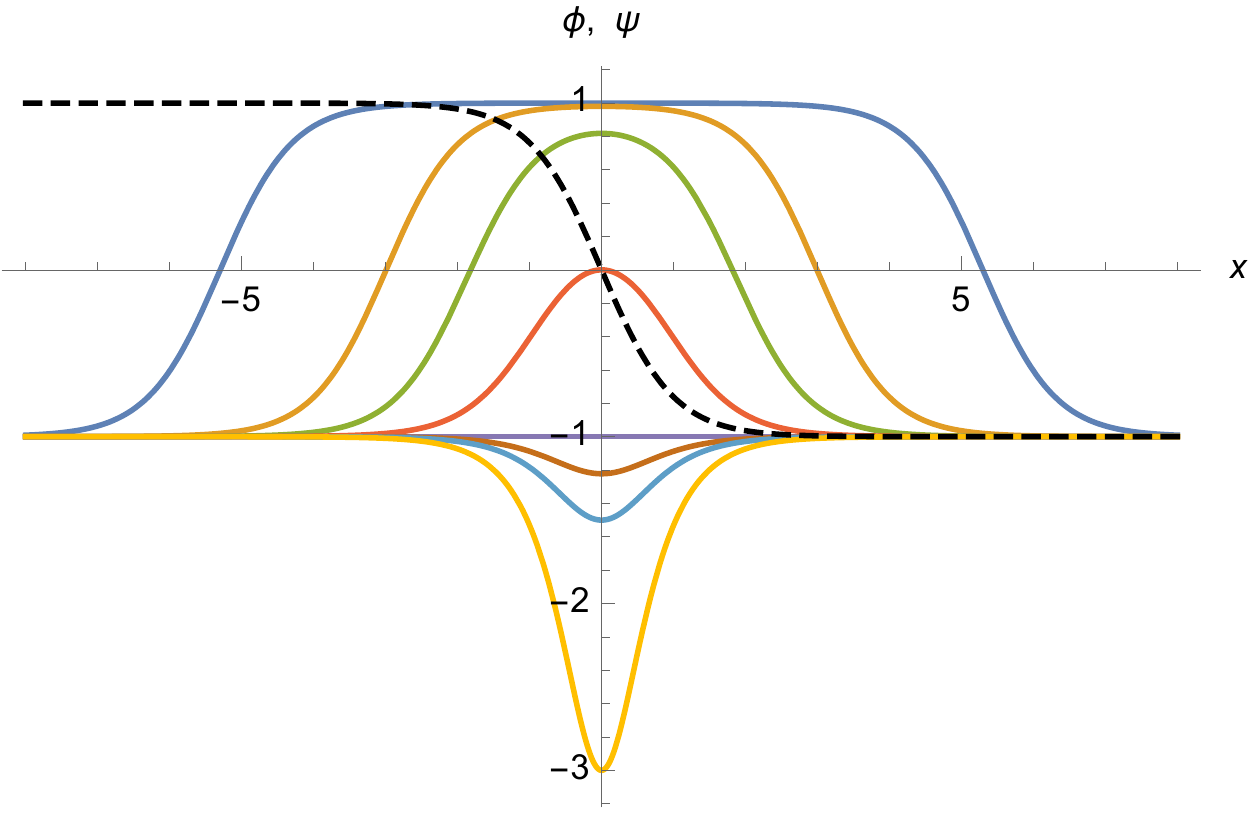}
\caption{Solutions in the nontrivial sector: dashed line - $\phi_0(x;a=0)=-\tanh(x)$; solid lines - $\psi_0(x;a=0,b)$ for several values of $b$.}
\label{sol-one}
\end{figure}

The topologically trivial (vacuum) sector, where $Q_\phi=0$, consists of a vacuum solution in the $\phi$ field and either a vacuum solution or a kink/antikink in the $\psi$ field. Namely,
\be
\phi_0^0 = \pm 1, \;\;\; \psi_0^0=\pm 1   \;\;\; \;\;\; \mbox{any sign combination}
\ee
for the pure vacuum field configurations, and
\be
\phi_0^0 =  1, \;\;\; \psi_0^0 =  \tanh(x-a)  \;\;\; \;\;\; \mbox{or} \;\;\; \;\;\; \phi_0^0 =  -1, \;\;\; \psi_0^0 =  -\tanh(x-a)
\label{Q0}
\ee
in the vacuum plus kink sector.
Note that the $\psi$ field carries a zero or non-zero value of the corresponding topological charge $Q_\psi \in \{ 0,\pm 1\}$. Due to the chirality of the Bogomol'nyi equations, there is only one solution with $Q_\psi=1$ and one with $Q_\psi=-1$. In other words, not all combinations are allowed. 

In the topologically nontrivial sector, $Q_\phi=\pm 1$, we find the $\phi^4$ kink or antikink. On the other hand, the $\psi$ field has a BPS solution carrying zero topological charge $Q_\psi$
\be
\phi_0^\pm = \pm \tanh(x-a), \;\;\; \psi_0^\pm = \mp \frac{b-\cosh^2(x-a)}{b+\cosh^2(x-a)}, \label{Q1}
\ee
where $a, b$ are two parameters (moduli) of the energetically equivalent solutions. In the subsequent analysis, we take $\phi_0^-=-\tanh(x-a)$ and the corresponding $\psi_0^-$ solution. The BPS solution of the $\psi$ field describes a pair of an infinitely separated $\tanh$ kink and antikink for $b\to \infty$,  which approach each other as $b$ decreases. At $b = 0$, the solitons completely annihilate to the $\psi=-1$ vacuum. As $b$ further decreases, the solution develops a negative bump whose bottom grows to an arbitrary negative value as $b \to -1$. Solutions for some values of $b$ are plotted in Fig. \ref{sol-one}.
\section{Moduli space}
The full dynamics of the model is governed by the Euler-Lagrange equations
\bea
& &-\phi_{tt}+\phi_{xx}+2\phi(1-\phi^2) -\phi(1-\psi^2)^2+(1-\psi^2) \psi_x=0, \label{eom1} \\
& & -\psi_{tt}+\psi_{xx} + 2\phi^2 \psi (1-\psi^2) - (1-\psi^2)\phi_x=0. \label{eom2}
\eea
However, it is well known that a small velocity interaction of BPS solitons, i.e., solutions of Bogomol'nyi equations, can be quite well approximated by the dynamics of a finite number of degrees of freedom, i.e., the {\it moduli}. This is due to the fact that the small velocity dynamics occurs via a sequence of energetically equivalent BPS solutions which saturate the topological energy bound in a fixed topological sector and which are labeled by a finite set of parameters $a^i, i=1...N$ called {\it moduli}. In a natural way they give rise to the {\it canonical moduli space}. This is a concept which arises if we want to reduce the full field theory dynamics, which has infinitely many degrees of freedom, to an approximation with a finite number of degrees of freedom. Then, the full space of field configurations $\mathcal{F}=\{ \Phi(x) \}$ is reduced to a set of particular configurations, $\mathcal{F}[a^i]=\{ \Phi_0(x; a^i)\}$, which should take into account the crucial properties of the field and correctly model its true time dynamics, at least in the small velocity limit. Here $\Phi=(\phi,\psi)$. 

The approximated dynamics is then given by the effective Lagrangian 
\be
L[a]=\frac{1}{2} g_{ij} \dot{a}^i \dot{a}^j - V(a)
\ee
obtained by the insertion of the restricted field configurations into the original Lagrangian. Here $g_{ij}$ are the moduli space metric components
\be
g_{ij}=\int_{-\infty}^{\infty} \frac{\partial \Phi_0}{\partial a^i} \frac{\partial \Phi_0}{\partial a^j}   dx
\ee
and $V(a)$ is the effective potential. Obviously, if the restricted space of fields consists of the BPS solutions, the potential is just a constant. In our case, the moduli space is canonically defined by the solutions of the Bogomolny equations.
 
In the topologically trivial sector, $Q_\phi=0$, there is only one modulus $a$, eq. (\ref{Q0}). The resulting one-dimensional metric and the effective potential are constants and take the form
\be
g_{aa}=\frac{4}{3}, \;\;\; V(a)=0.
\ee
The dynamics is trivial and describes the (anti)kink moving with a constant velocity i.e., $\psi=\pm \tanh(x - vt)$.

\begin{figure}
\includegraphics[width=0.6\textwidth]{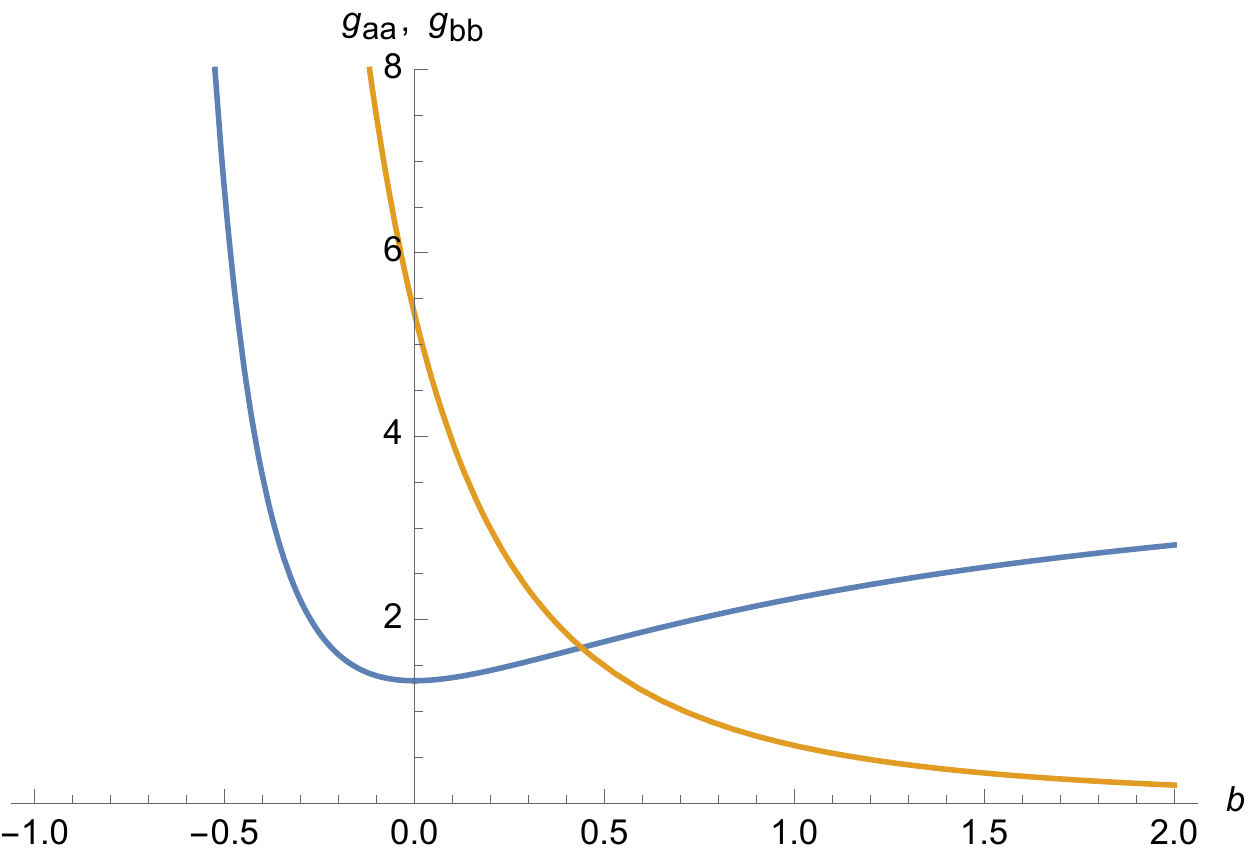}
\caption{Moduli space metric components in $Q_\phi=\pm 1$ BPS sector: $g_{aa}$ (blue) and $g_{bb}$ (orange).}
\label{metric-plot}
\end{figure}

From now on, we will consider the more interesting, topologically nontrivial sector, $Q_\phi=\pm 1$, where we have found two moduli $a$ and $b$, eq. (\ref{Q1}). Then we find that $L[a,b]$ is given by
\be
L[a,b]=\frac{1}{2} g_{aa} (b) \dot{a}^2 + \frac{1}{2} g_{bb} (b) \dot{b}^2 - \frac{4}{3}, 
\ee
where the metric components are (see Fig. \ref{metric-plot})
\be
g_{aa}(b)=\frac{4}{3} + \frac{2}{3(1+b)^2} \left( 3+4b(1+b) - \frac{3(1+2b)}{\sqrt{b}\sqrt{1+b}} \, \mbox{arctanh} \sqrt{\frac{b}{1+b}} \right), \label{gaa}
\ee
\be
g_{bb}(b)=\frac{1}{6b(1+b)^3} \left( -3+4b(4+b) + \frac{3(1+6b)}{\sqrt{b}\sqrt{1+b}} \, \mbox{arctanh} \sqrt{\frac{b}{1+b}} \right) \label{gbb}
\ee
and $g_{ab}=0$.  Asymptotically, for $b\to \infty$, which corresponds to an infinitely separated pair of solitons of the $\psi$ field, the components tend to finite values. Specifically, $g_{bb}(\infty)=0$ and $g_{aa}(\infty)=\frac{12}{3}$ (i.e., three times the mass of the single kink). At $b=0$, where the initial solitons in the $\psi$ field annihilate to the $-1$ vacuum, the metric components are $g_{bb}(0)=\frac{16}{3}$, $g_{aa}(0)=\frac{4}{3}$. Finally, for $b \to -1$ the metric diverges, $g_{aa}(-1)=g_{bb}(-1) =\infty$. One can show that the moduli space is metrically complete. Indeed, the boundaries $b \to -1$ and $b\to \infty$ are approached in infinite time as the corresponding field configurations have infinite distance from any other BPS solution. Here the distance between two sets of fields $(\phi_1,\psi_1)$ and $(\phi_2,\psi_2)$ is computed by using the natural, infinite-dimensional Euclidean metric 
\be
s^2=\int_{-\infty}^\infty \left( \phi_1-\phi_2 \right)^2+\left( \psi_1-\psi_2\right)^2 dx.
\ee
The resulting equations of motion of the moduli are
\bea
\frac{d}{dt} (g_{aa}\dot{a} )=0, \\
\frac{d}{dt} (g_{bb}\dot{b} ) - \frac{1}{2} g_{bb}' \dot{b}^2-\frac{1}{2} g_{aa}' \dot{a}^2=0,
\eea
where dot and prime denote differentiation w.r.t. time and $b$ respectively. This set of second order equations integrates to two first order equations
\bea
P&=&g_{aa}\dot{a}, \\
E&=&\frac{1}{2} g_{bb} \dot{b}^2+\frac{1}{2} g_{aa} \dot{a}^2 +\frac{4}{3},
\eea
where $P$ and $E$ are conserved quantities, i.e., momentum and total energy. They can be combined and lead to
\be
\frac{db}{dt} = \pm \sqrt{\frac{1}{g_{bb}} \left( 2E-\frac{8}{3} - \frac{P^2}{g_{aa}}\right)}.
\ee
There is also a solution with $a=a_0=\mbox{const.}$, which corresponds to a non-moving kink of the $\phi^4$ field. Then, the moduli space dynamics describes a collision of a kink-antikink pair of the $\psi$ field.

\section{Spectral structure}
The canonical moduli space (\ref{gaa}), (\ref{gbb}) describes dynamics of the kinetic degrees of freedom corresponding to zero energy excitations, i.e., {\it zero modes} of the full model. Indeed, as all (BPS) configurations parametrized by the moduli coordinates $a,b$ have the same energy, a transition from one configuration to another one costs an arbitrarily small amount of energy (for arbitrarily small velocities). In the next step, we should obviously include non-zero mass excitations of the fields, that is, the massive vibrations, i.e., {\it normal (bound) modes} $\zeta_k(x;a^i)$, where $k=1..M$ counts the modes. Then, the reduced field space is 
\be
\mathcal{F}[a^i,A^k]=\{\Phi_0(x;a^i)+\sum_{k=1}^M  A^k\zeta_k(x;a^i)\}
\ee
and describes the excitations of the BPS solitons.  Now the dynamics of the fields is defined by a bigger moduli space, usually referred to as the {\it vibrational moduli space}. It should be stressed that, in general, the spectral structure, i.e., the number and the form of the normal modes as well as their frequencies, may nontrivially depend on the moduli $a^i$ parametrizing the BPS solutions. This is the case considered in the present work. Physically it is a situation where the mode structure varies as the solitons approach each other. To underline such a dependence we call this vibrational moduli space the {\it dynamical vibrational moduli space}. This should be contrasted with the vibrational moduli space considered in the typical multi-soliton collisions like, e.g., the kink-antikink collisions in the $\phi^4$ model, where the normal modes added on top of the scattered solitons are {\it frozen} in the form derived in the free kink (antikink) limit. Thus, this vibrational space of configurations should be called the {\it frozen vibrational moduli space}. Such a construction is widely applied to non-BPS processes like kink-antikink annihilations. Here we show that, at least in the case of excited BPS processes, the dynamical vibrational moduli space is more adequate to reproduce the dynamics of topological solitons. 

Furthermore, contrary to the canonical moduli space, the dynamical vibrational moduli space may be metrically incomplete, which means that during the evolution we can reach a point where the corresponding metric has a singularity. However, this is not just an apparent singularity, i.e., merely a mathematical consequence of a badly chosen coordinate in the vibrational moduli space which could be cured by a suitable redefinition of the coordinates (conf. the zero vector problem in the $\phi^4$ theory \cite{moduli-we}). This is a genuine singularity resulting from the dynamical (changeable) nature of the normal modes, which occurs as a consequence of the transition (disappearance) of a normal mode into the continuous spectrum. Indeed, for a finite value of one of the moduli $a^i$ (related to the distance between the solitons), one of the additional coordinates $A^k$ parametrizing the vibrational space  simply does not exist any longer. Therefore, the existence of such a point does not disqualify the vibrational moduli space. On the contrary, it has physical importance as it gives rise to a thin spectral wall. 

Let us now discuss the flow of the normal modes in the case of the $Q_\phi=-1$ sector.  So, let us consider small perturbations of the BPS solutions $\phi(x,t)=\phi_0^-+\eta(x,t)$ and $\psi (x,t) =\psi_0^- + \xi (x,t)$.  (Here, $\zeta=(\eta,\xi)$). The linear perturbation equations then take the form (from now on we omit the superscript $(-)$ in the BPS solutions)
\bea
-\eta_{tt}+\eta_{xx} + \left[ 2(1-3\phi_0^2)-(1-\psi_0^2)^2\right] \eta -2 \left[ \psi_0 \psi_{0,x}-2\phi_0\psi_0(1-\psi_0^2) \right]\xi + (1-\psi_0^2)\xi_x&=&0,
\\ -\xi_{tt}+\xi_{xx} +2\left[ \phi_0^2(1-3\psi_0^2)+\psi_0\phi_{0,x}\right]\xi +4\phi_0\psi_0(1-\psi_0^2)\eta -(1-\psi_0^2)\eta_x&=&0.
 \eea
 Assuming that $\eta(x,t)= \eta(x) \sin \omega t $, $\xi (x,t)= \xi (x) \sin \omega t$ and using the Bogomol'nyi equations $\psi_{0,x}=\phi_0(1-\psi_0^2)$ we get
 \bea
\eta_{xx} + \left[ 2(1-3\phi_0^2)-(1-\psi_0^2)^2\right] \eta +2\psi_0\psi_{0,x} \xi +(1-\psi_0^2)\xi_x&=&-\omega^2 \eta, \label{self1}
\\
\xi_{xx} +2\left[ \phi_0^2(1-3\psi_0^2)+\psi_0\phi_{0,x}\right]\xi +4\psi_0\psi_{0,x} \eta -(1-\psi_0^2)\eta_x&=& - \omega^2 \xi. \label{self2}
 \eea
 Note that the $a$ dependence of the spectral problem is trivial. Indeed, the modulus  $a$ can be removed by a coordinate redefinition $x \to x +a$. Therefore, the normal modes (and frequencies) are only functions of the $b$ modulus, $\omega(b)$.
 
 Before we compute the flow of the normal modes on the canonical moduli space, that is, their change as we vary $b$, let us consider the asymptotic case $b\to \infty$, when the BPS solution consists of three infinitely separated solitons: left $(l)$, central $(c)$ and right $(r)$
 \be
 \phi_0^l=1, \psi_0^l=\tanh(x+c), \;\;\; \phi_0^c=- \tanh x, \psi_0^c=1, \;\;\; \phi_0^r=-1, \psi_0^r=-\tanh(x-c),
 \ee
  each leading to its own spectral structure with the infinitely separated eigenmodes. Here $c \to \infty$. First of all, the central soliton, localized at the origin, gives rise to a very simple spectral problem
   \bea
\eta_{xx} - \left(4-\frac{6}{\cosh^2x} \right) \eta  &=&-\omega^2 \eta, 
\\
\xi_{xx} -\left( 4-\frac{2}{\cosh^2x}\right) \xi &=& - \omega^2 \xi.
 \eea
 \begin{figure}
\includegraphics[width=1.0\textwidth]{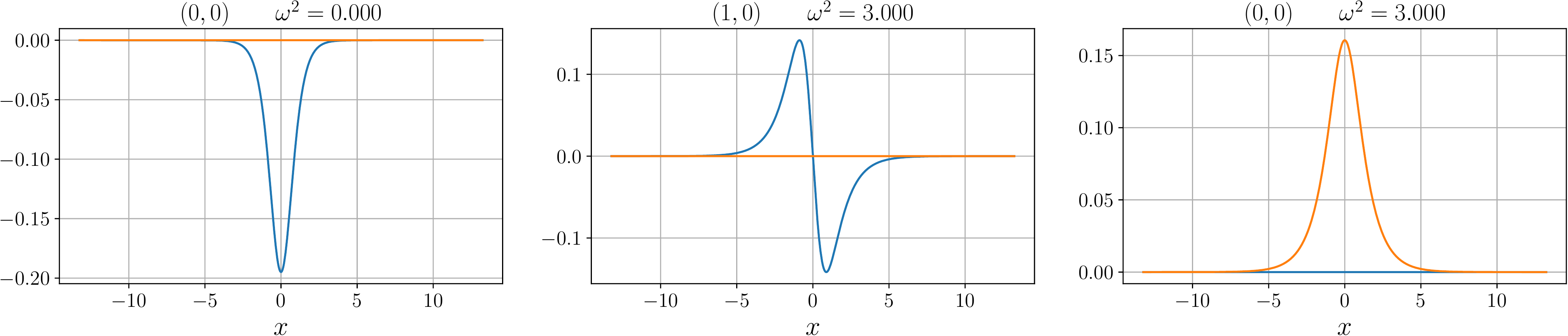}
\caption{Eigenmodes $\eta, \xi$ (blue, orange) for $b\to \infty$ - the central soliton $\phi_0^c=- \tanh x, \psi_0^c=1$.}
\label{mode-c}
\vspace*{0.4cm}
\includegraphics[width=1.0\textwidth]{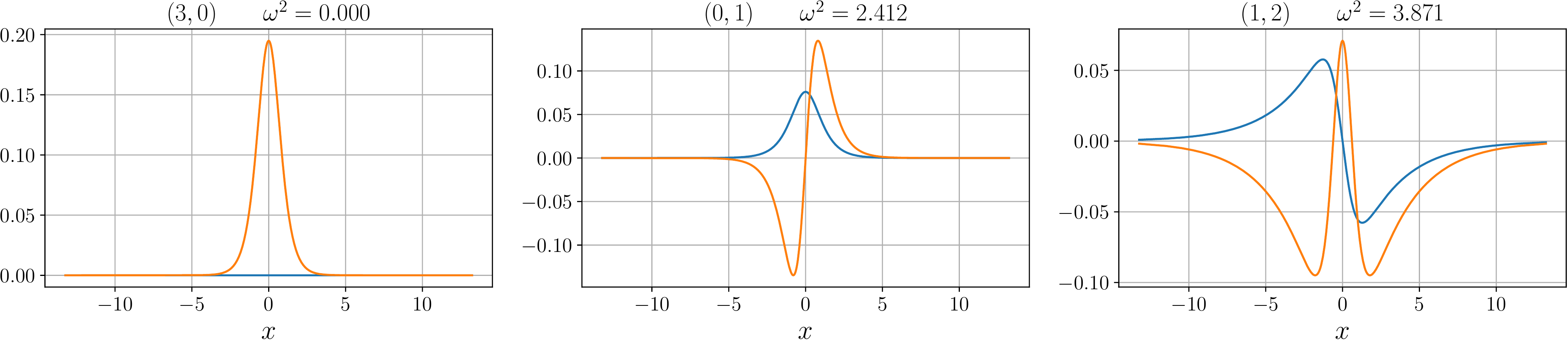}
\caption{Eigenmodes $\eta, \xi$ (blue, orange) for $b\to \infty$ - the left soliton $\phi_0^l=1, \psi_0^l=\tanh x$.}
\label{mode-l}
\end{figure}
 The effective potential for $\eta$ is identical to the $\phi^4$ case and leads to a zero-mode and a vibrational mode with $\omega^2=3$ being exactly the usual shape mode 
 \be
 \eta=\frac{\sinh x}{ \cosh^2 x}.
 \ee  
 The effective potential for $\xi$ is the effective potential in the spectral problem for the sine-Gordon soliton, shifted by 3. Hence the original zero mode is shifted to $\omega^2=3$. The corresponding eigenfunction is 
 \be
 \xi=\frac{1}{\cosh x}.
 \ee 
Thus, the central soliton carries one zero and two degenerate vibrational modes, see Fig. \ref{mode-c}. 
 Next, the spectral problem for the left (right) soliton does not separate due to non-zero off-diagonal as well as first derivative terms. As a consequence, the mode structure does not simply coincide with the $\phi^4$ theory case. On the contrary, the left (as well as right) soliton gives rise to one zero mode and two vibrational modes, Fig. \ref{mode-l}. Surprisingly, these eigenmodes can be found in an analytical form (for simplicity we shift $c$ to 0)
\bea
\omega^2_p&=&4-p^2, \;\;\; \eta_p=\frac{p-1}{\cosh^p x}, \;\;\; \xi_p = \frac{ \sinh x}{\cosh^{p+1} x }, \;\;\; p=2^{1/3},\\
\omega^2_q&=&3 + 2 q - q^2, \;\;\; \eta_q = \frac{\sinh x}{\cosh^q x}, \;\;\;  \xi_q = \frac{1}{q-1} \frac{1}{\cosh^{q+1}x} \left( \frac{q+1}{q+2} \cosh^2 x -1 \right),
\eea 
where $q$ is a root of the following algebraic equation: $-3 - q + q^2 + q^3 = 0$, given by
\be
q=\frac{1}{3} \left(-1 + \left(35 - 3 \sqrt{129} \right)^{1/3} + \left(35 + 3 \sqrt{129}\right)^{1/3}\right)  \approx 1.359.
\ee
Numerical values are $\omega^2_p\approx 2.412$ and $\omega^2_q \approx 3.871$.
Altogether, for $b \to \infty$ we have three zero modes and six vibrational modes, see Fig. \ref{spectral}. As $b$ decreases, the kink and antikink in the $\psi$ field come closer to the soliton in the $\phi$ field and the modes of the asymptotic states (infinitely separated solitons) interact and mix. So there are still nine modes for sufficiently large, finite $b$. For simplicity we plot them in colors:
\begin{itemize}
\item Two (black) zero modes ($n=0,1$) corresponding to the two moduli $a$ and $b$. One represents the simultaneous translation of {\it all} solitons $x \to x +a$. The second is the symmetric superposition of the zero modes of the asymptotic (infinitely separated) kink and antikinks in the $\psi$ field and represents their flow as dictated by the BPS solution. 
 \begin{figure}
\includegraphics[width=1.0\textwidth]{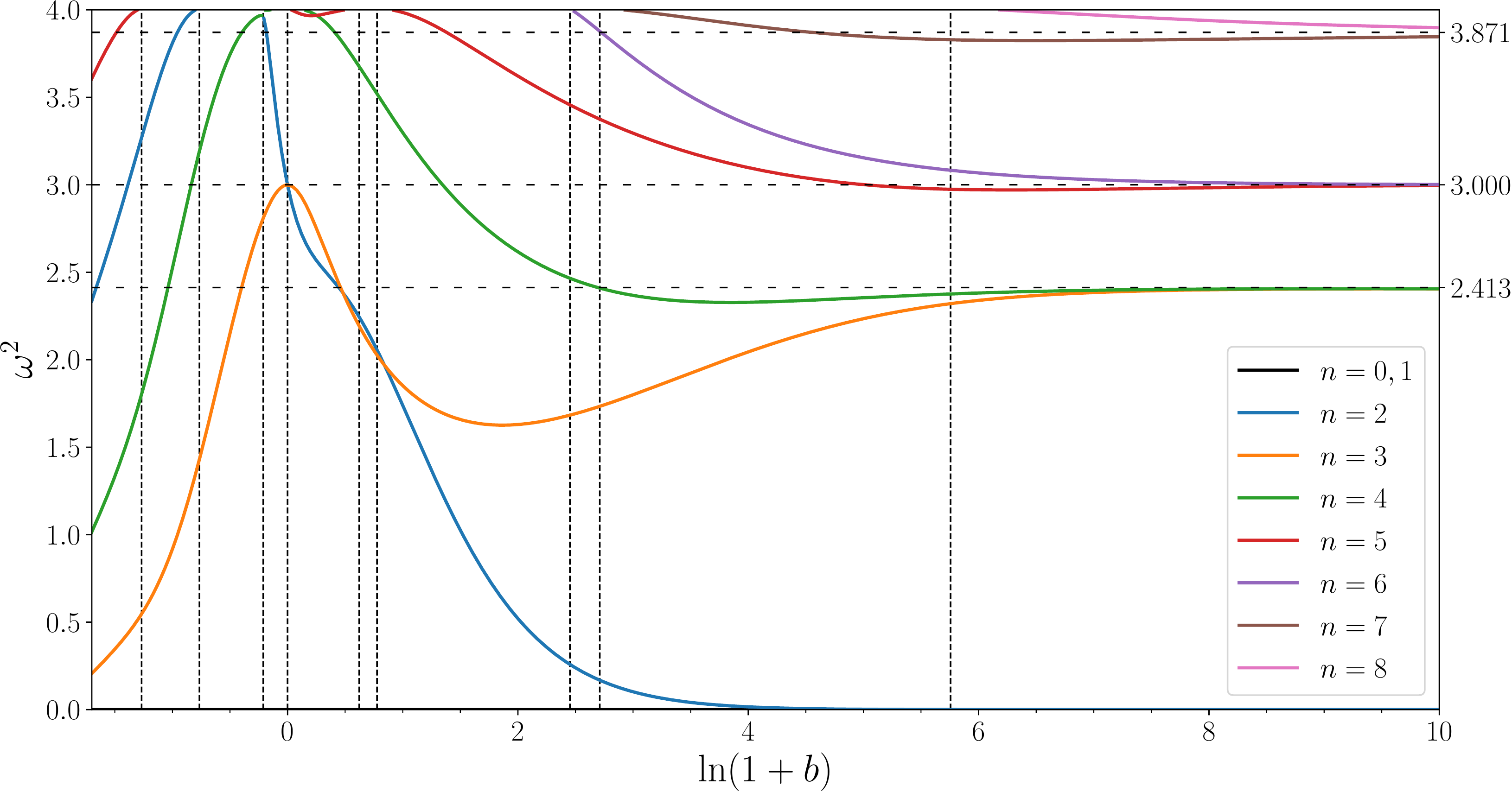}
\caption{The flow of the spectral structure on the moduli space.}
\label{spectral}
\end{figure}
\item The blue mode $(n=2)$. This is initially a very deep mode with the frequency arbitrarily close to 0 as $b$ grows. It is built out of the antisymmetric superposition of the zero modes of the asymptotic (infinitely separated) kink and antikink in the $\psi$ field. Hence, it may be excited by a simultaneous shift of the position of the kink and antikink by the same constant $\Delta x_0$. It crosses the mass threshold at $b=-0.53$ and at $b=-0.19$.
\item The orange and green modes ($n=3$ and $n=4$) which are superpositions of the massive $p$ modes of the asymptotic (infinitely separated) kink and antikinks in the $\psi$ field. Specifically, the orange mode is formed by a symmetric superposition of the $p$ modes while the green mode represents an antisymmetric superposition. The orange mode doesn't cross the mass threshold, while the green mode only touches it at $b=0$.
\item The red and purple modes ($n=5$ and $n=6$) are the modes of the central antikink in the $\phi$ field. The red mode originates in the mode where $\eta$ is the shape mode while the purple mode originates in the mode where $\xi=1/\cosh x$. Of course, as $b$ decreases the other components also appear. The purple mode crosses the mass threshold at $b=10.63$. The red mode does it a few times at $b=1.18, 0.87, 0, -0.74$.
\item The brown and pink modes ($n=7$ and $n=8$). They are the two most shallow modes which first enter the continuous spectrum. They are again superpositions of the massive $q$ modes of the asymptotic (infinitely separated) kink and antikink in the $\psi$ field. Specifically, the brown mode is formed by a symmetric superposition of the $q$ modes while the pink mode represents the antisymmetric superposition of the $q$ modes. They cross the mass threshold at $b=321$ (pink) and $b=14.11$ (brown).
\end{itemize}

In order to solve the coupled eigenvalue problem we used two independent methods which gave consistent results. 
In the first method we discretized the linearized equation on a spatial grid consisting of $N$ points ($N$ was between 50 and 500). 
We used equidistant points in the real space for $x\in[-L,L]$, where $L$ was a large number  (typically 10, 15 or 20), and in the 
compactified space defined by a new coordinate $y=2\arctan (x/L)/\pi$ (here L was smaller). To discretize the derivative we used the standard three point stencil. 
The Sturm-Liouville operator was represented as a large $2N\times2N$ sparse  matrix consisting of four $N\times N$  blocks, each of 
which was a tri-diagonal matrix. Next we applied a standard eigenvalue problem solver for sparse matrices using Python and Julia 
languages to find the lowest eigenvalues. This method gave reliable results for modes which were well below the threshold, while it led to significant errors if a mode approached the mass threshold. Therefore, the positions of the spectral walls were found with rather large errors.

To calculate more precisely the modes, especially near the threshold, we used a shooting method. For different values of $\omega$ we solved the linearized equation (\ref{self1})-(\ref{self2}) using a standard ODE initial value problem starting at $x=0$ and assuming certain symmetry $(\xi,\eta,\xi',\eta')=(\cos s, 0, 0, \sin s)$ or $(0, \cos s, \sin s, 0)$ matching with the asymptotic form of the solutions $(\xi,\eta)\approx A e^{-kx}+Be^{kx}$ for $x>L$. To find the eigensolutions we required $B=(0,0)$. This method had two shooting parameters $s$ and $\omega<2$. Again, finding an interesting eigenmode near the threshold was still a challenging problem. However, it was more reliable than the first method. Note that in order to find the position of a spectral wall we have simply required $\omega=2$ and we needed to find $s$ and $b$. In this case $b$ was not constrained. This allowed much more accurate calculations.

\vspace*{0.1cm}

For $b=0$ we can also determine the mode structure in an analytical way. The solution for the $\psi$ field is just the $\psi_0=-1$ vacuum. Then, the spectral problem greatly simplifies
 \bea
\eta_{xx} - \left(4-\frac{6}{\cosh^2x} \right) \eta  &=&-\omega^2 \eta, 
\\
\xi_{xx} -\left(4-\frac{6}{\cosh^2x} \right) \xi &=& - \omega^2 \xi.
 \eea
Both effective potentials coincide and they have become the P\"{o}schl-Teller potential of the $\phi^4$ kink. Therefore, there are two zero modes (corresponding to the flow of the moduli $a$ and $b$) and two vibrational modes with $\omega^2=3$. This shows that five modes had to cross the mass threshold as we change $b$ from $\infty$ to $0$. Note that, although $\omega^2=3$, these vibrational modes do not originate from the modes of the central solitons for $b\to \infty$. Instead, they come from one of the zero modes and one of the modes with $\omega^2_p$, see Fig. \ref{spectral}. 

There are several significant characteristic features of the spectral structure as we move in the moduli space, i.e., change $b$ from infinity to -1, see Fig. \ref{spectral}. First of all, as we have already said, vibrational modes can cross the mass threshold and enter into the continuous spectrum. The corresponding points of the moduli space $b_{sw}$, where the mode structure admits a non-continuous change, are important as they are responsible for thin spectral walls. Secondly, a new vibrational mode can appear from the continuous spectrum. Again, this results in a thin spectral wall. Interestingly, some modes cross each other. This is a new property not observed in the soliton dynamics of the BPS impurity model.  However, as we focus on the spectral walls, the impact of such a mode crossing on the soliton dynamics has not been studied in the current work.  

We have stopped the analysis of the spectral structure at $\ln(1+b)\approx -2$ due to the large numerical errors. However, this regime is far beyond the regime of $b$ necessary for the observation of thin spectral walls. We would also like to remark that the computation of the positions of the crossing of the mass threshold is a very difficult numerical task, especially if a mode enters the continuum slowly. Therefore, there can be some uncertainties in the numerically found values.

\section{Thin spectral walls}
To find a thin spectral wall we have to excite one of the modes in the asymptotic state which consists of initially separated solitons of the $\psi$ field ($b\to \infty$) and the antikink of the $\phi$ field. Then, we boost the solitons of the $\psi$ field towards each other. This obviously corresponds to a flow in the moduli space with decreasing $b$ while $a$ is kept at zero. There are seven modes which can be excited this way. A natural choice is to excite a mode which enters the continuum spectrum for relatively well separated $\phi$ solitons (larger values of $b$). Here we choose the pink and purple modes. In the case of the pink mode ($n=8$) the initial configuration for our numerical computation is given by
\bea
\phi_{in}&=& -\tanh x +A_1 \eta_q (x+x_0-vt) \sin (\omega_q t)-A_2 \eta_q (x-x_0+vt) \sin (\omega_q t), \label{init-phi} \\
\psi_{in}&=& \tanh (x+x_0-vt) +A_1 \xi_q (x+x_0-vt) \sin (\omega_q t) \nonumber
\\
&-&  \tanh(x-x_0+vt) - A_2 \xi_q (x-x_0+vt) \sin (\omega_q t) -1, \label{init-psi}
\eea
where $x_0$ is the initial distance between the neighbouring solitons, $v$ the initial velocity of the outer solitons and $A_1=-A_2$ is the amplitude of the exited mode. Then this can be evolved in the system of the full set of equations of motion (\ref{eom1})-(\ref{eom2}) using 4th order Runge-Kutta method from the GSL Library. In our numerics we have considered very small initial velocities, which allows us to assume  the nonrelativistic approximation in the expression for the initial state, i.e., $\gamma=(1-v^2)^{-1/2} \approx 1$. 
 \begin{figure}
 \hspace*{-0.8cm}\includegraphics[width=0.35\textwidth]{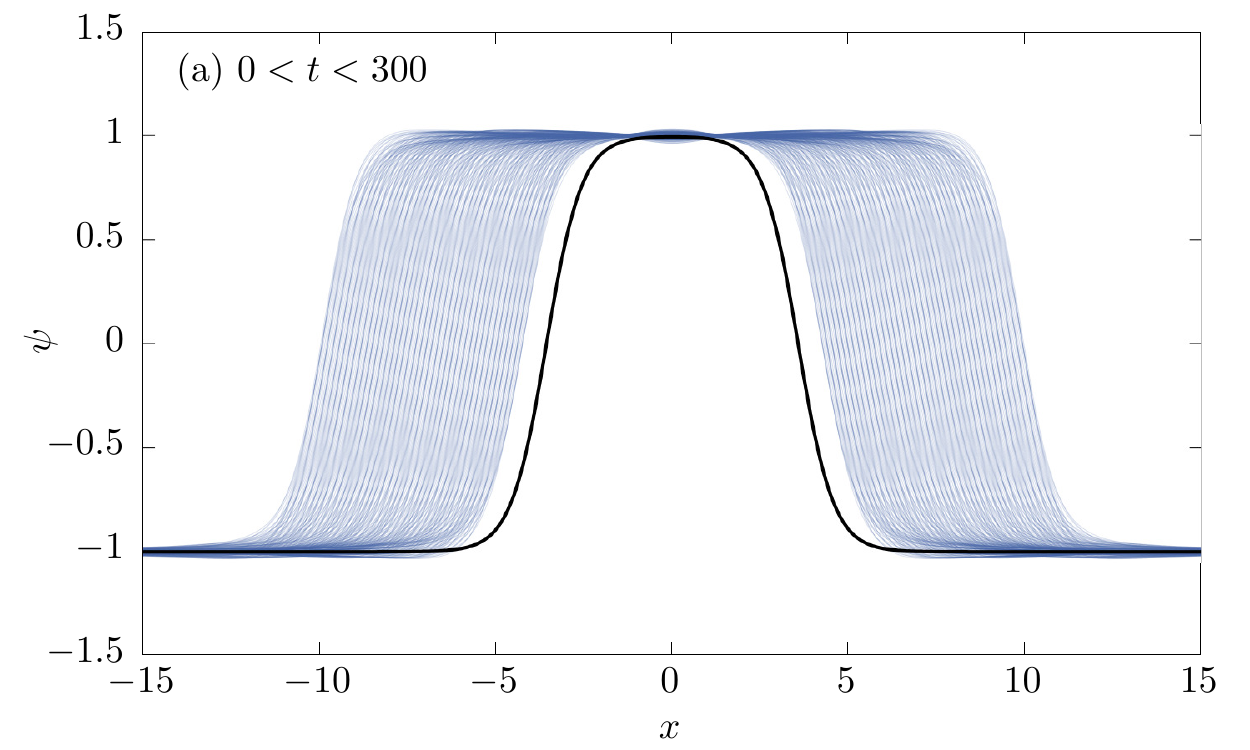}
\hspace*{-0.3cm} \includegraphics[width=0.35\textwidth]{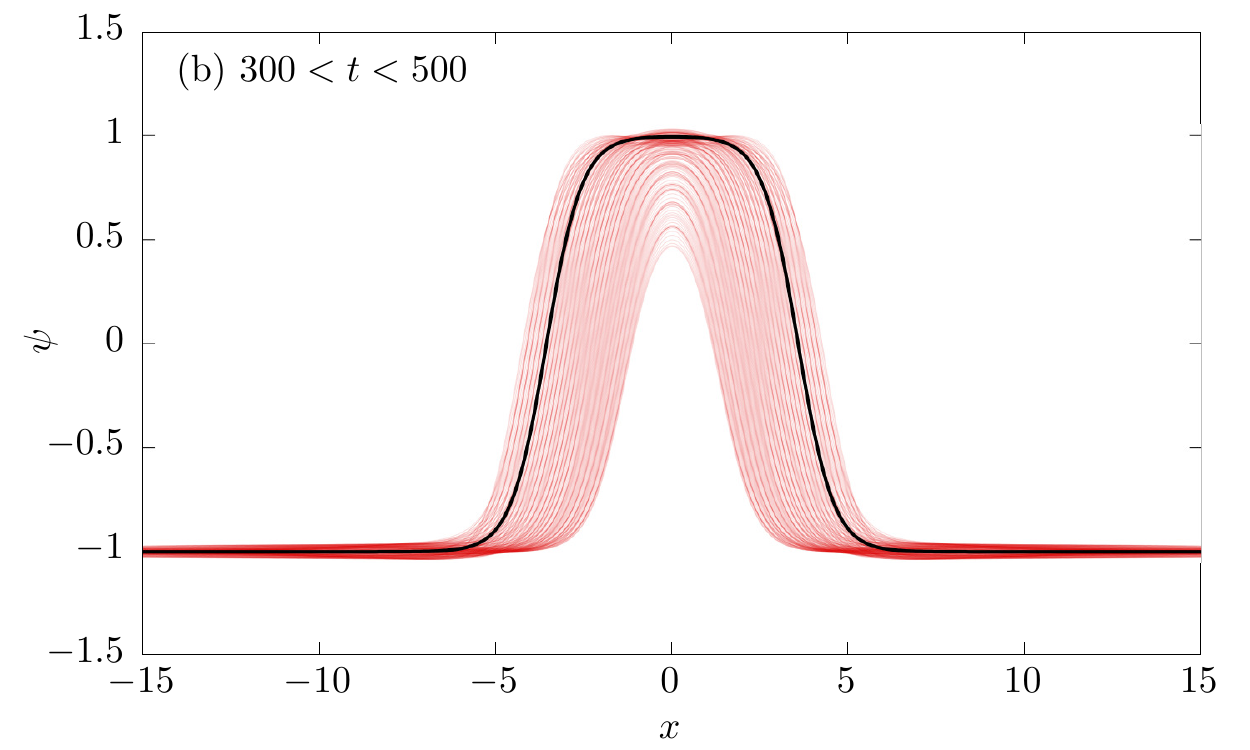}
\hspace*{-0.3cm} \includegraphics[width=0.35\textwidth]{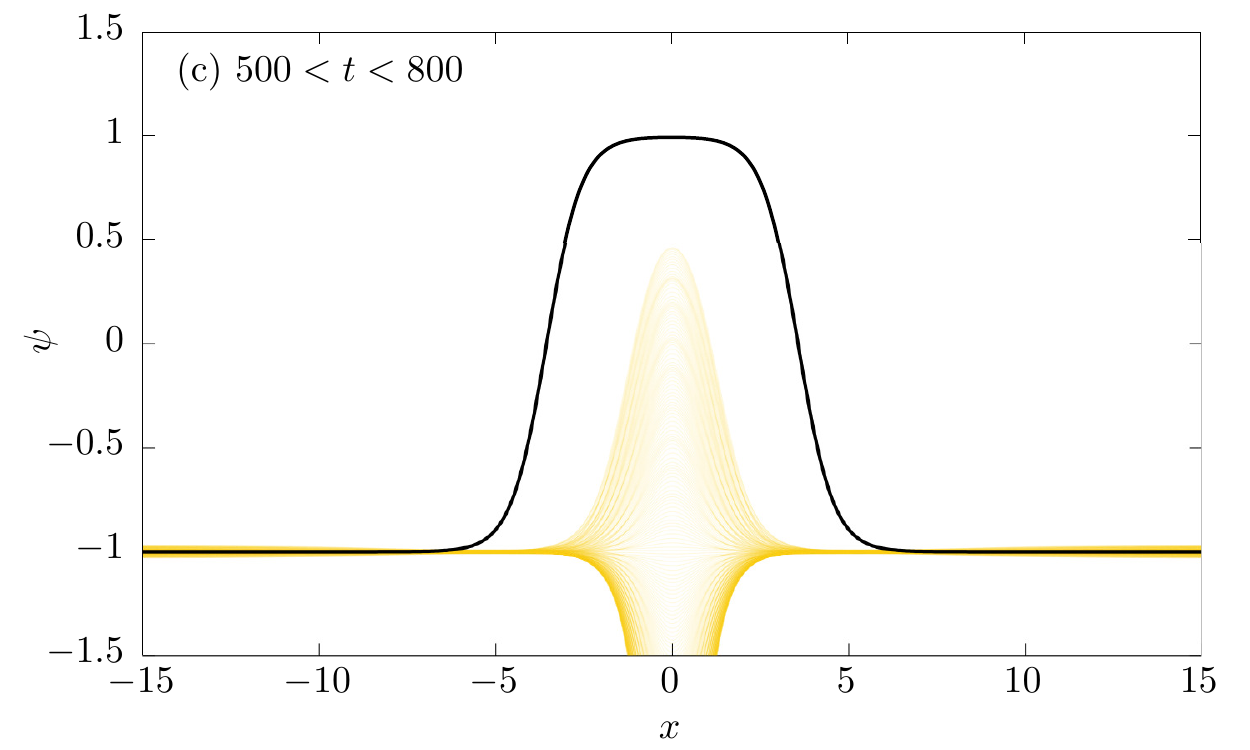}
\caption{Evolution of the field $\psi$ obtained in the scattering of the initial configuration (\ref{init-phi})-(\ref{init-psi}) with the pink mode $(n=8)$ initially excited. Here $A=0.04$ and solitons pass the SW. }
\label{pink-wall-1}
\vspace*{0.3cm}
\hspace*{-0.8cm}\includegraphics[width=0.35\textwidth]{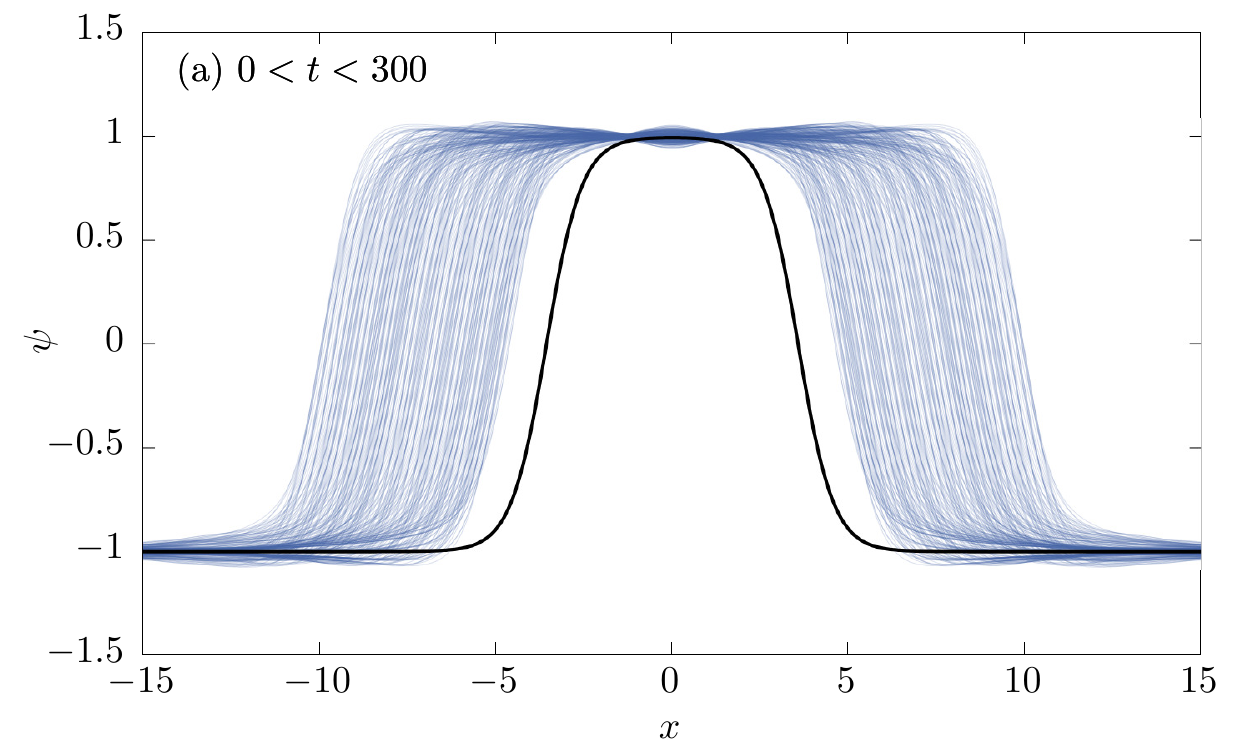}
\hspace*{-0.3cm} \includegraphics[width=0.35\textwidth]{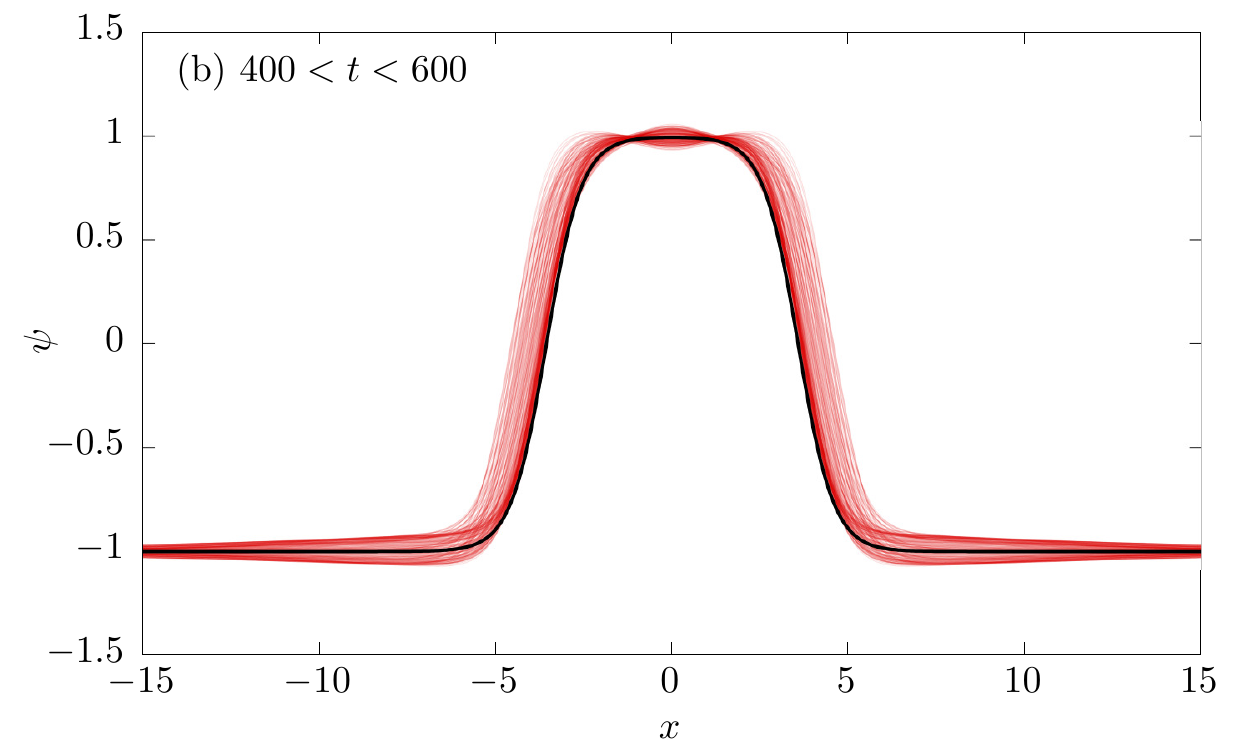}
\hspace*{-0.3cm} \includegraphics[width=0.35\textwidth]{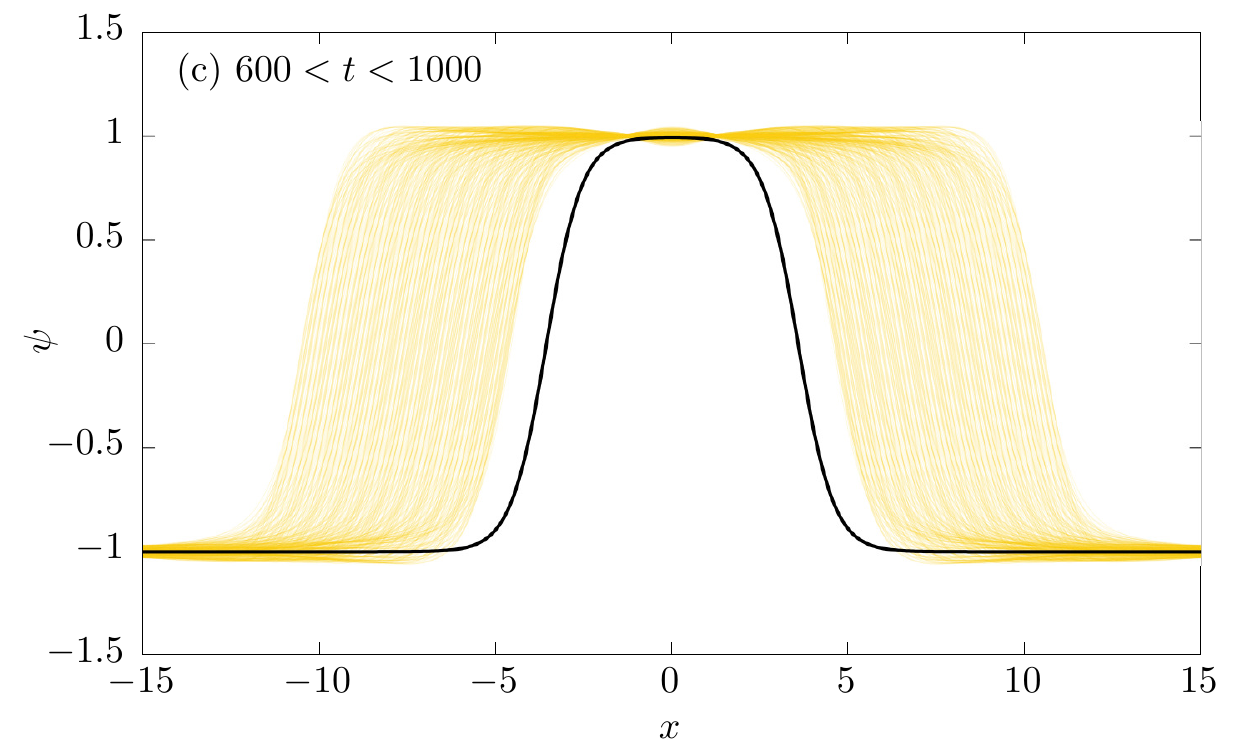}
\caption{Evolution of the field $\psi$ obtained in the scattering of the initial configuration (\ref{init-phi})-(\ref{init-psi}) with the pink mode $(n=8)$ initially excited. Here $A=0.078$ and solitons form a long living stationary solution ($t \in [400,700]$) localized at the BPS solution with $b=321$. }
\label{pink-wall-2}
\vspace*{0.3cm}
 \hspace*{-0.8cm}\includegraphics[width=0.35\textwidth]{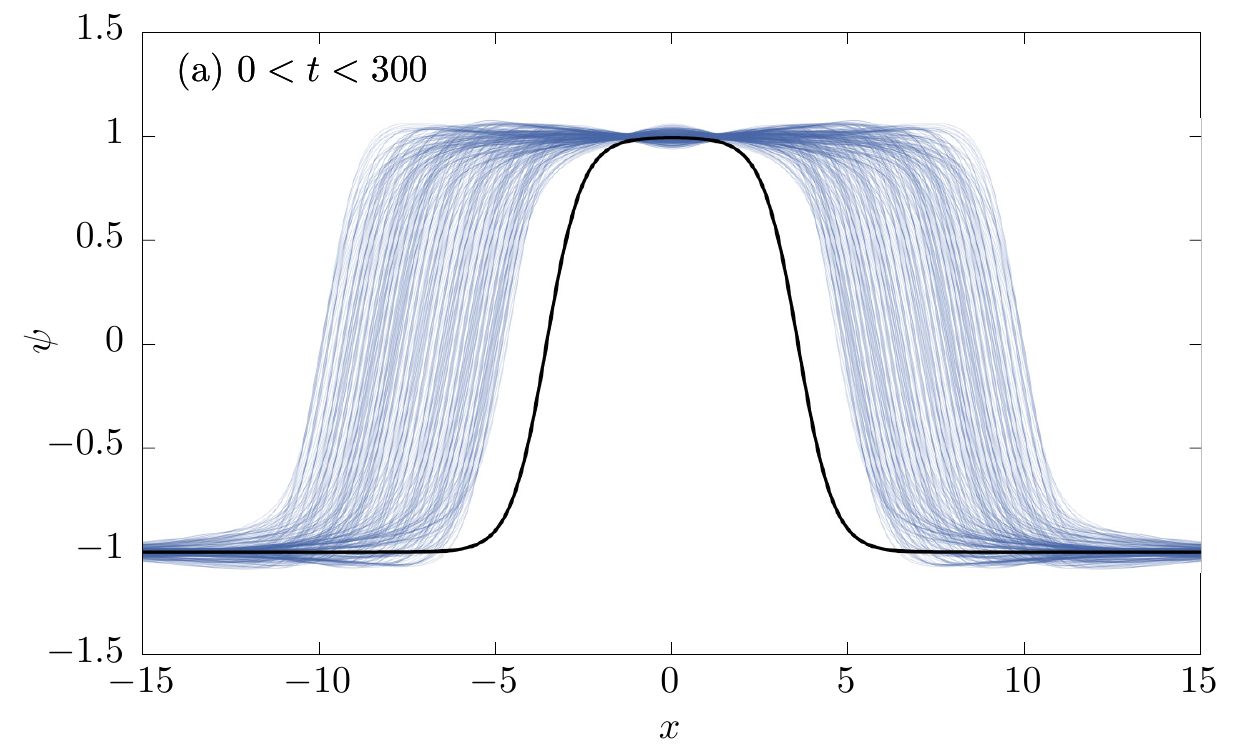}
\hspace*{-0.3cm} \includegraphics[width=0.35\textwidth]{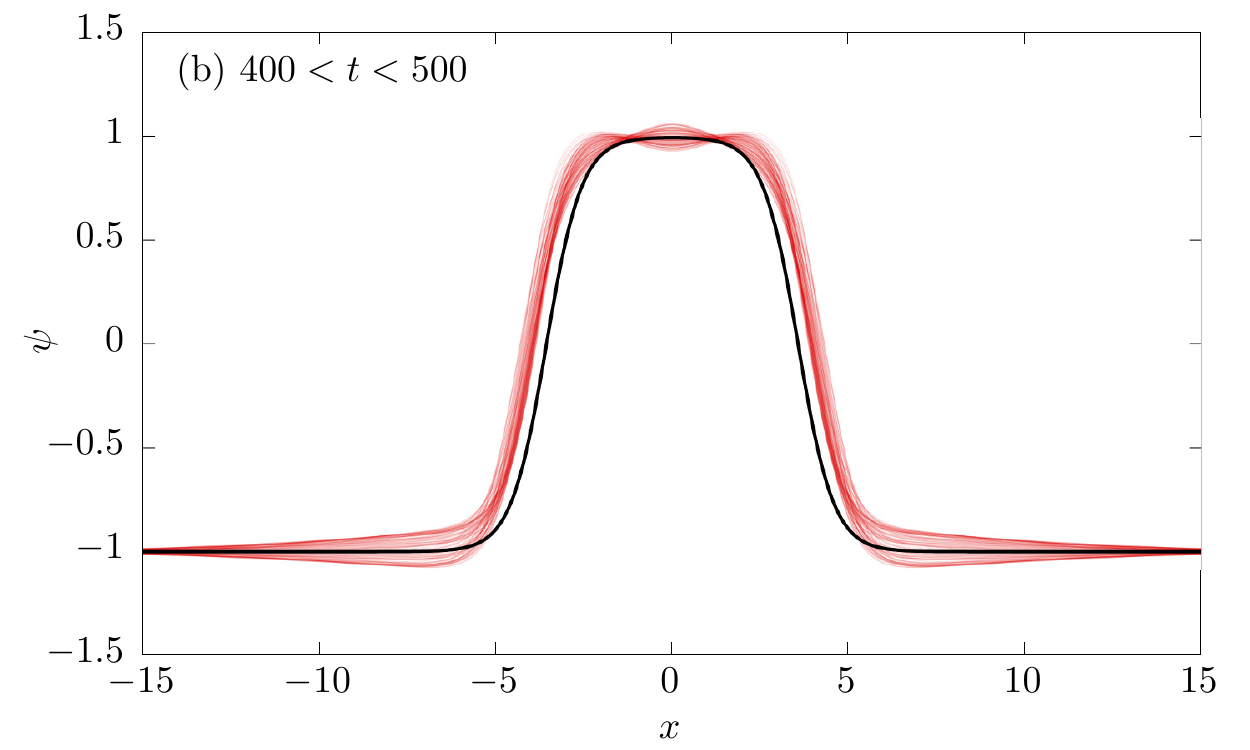}
\hspace*{-0.3cm} \includegraphics[width=0.35\textwidth]{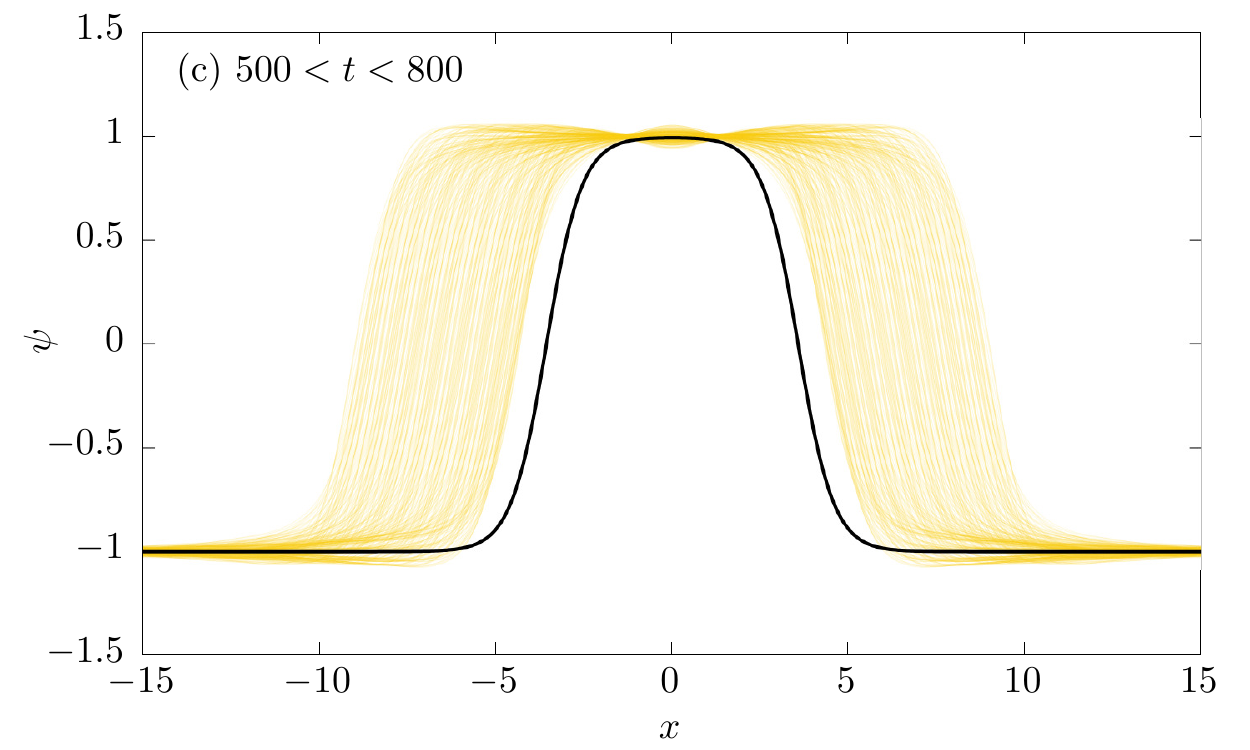}
\caption{Evolution of the field $\psi$ obtained in the scattering of the initial configuration (\ref{init-phi})-(\ref{init-psi}) with the pink mode $(n=8)$ initially excited. Here $A=0.085$ and solitons are back scattered before the SW.}
\label{pink-wall-3}
\end{figure}
\begin{figure}
 \hspace*{-0.8cm}\includegraphics[width=0.35\textwidth]{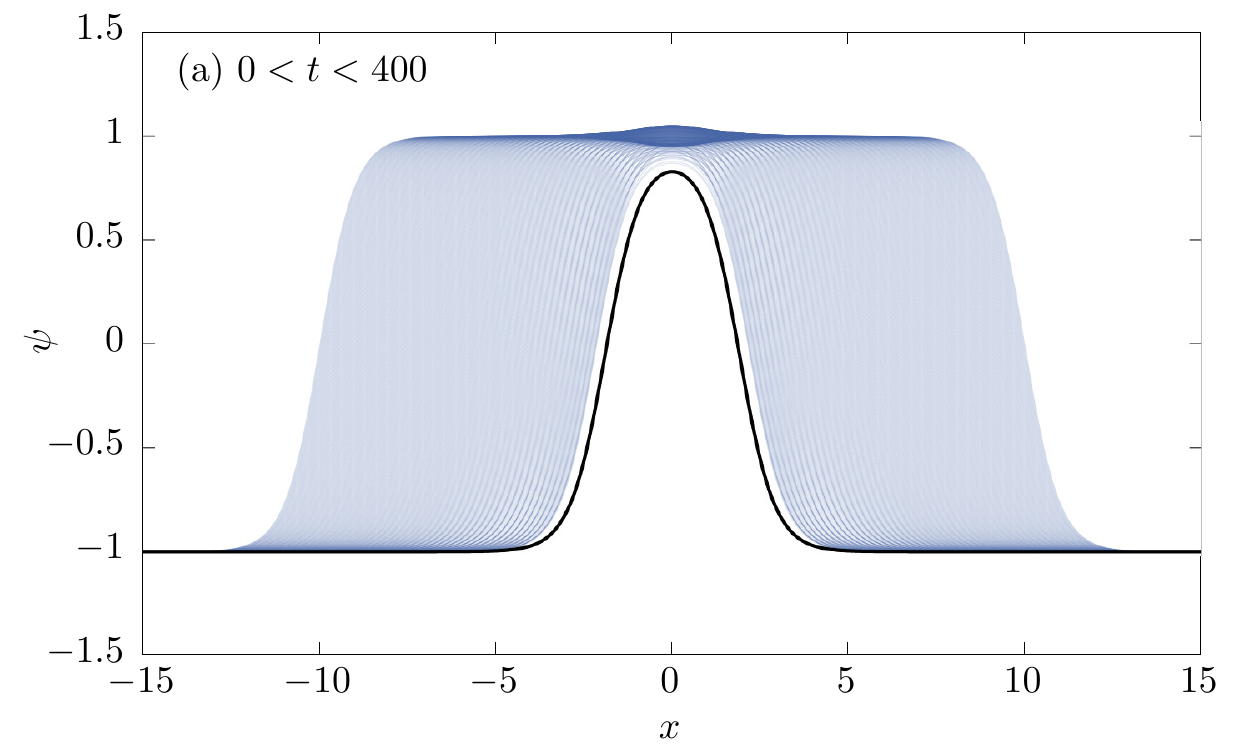}
\hspace*{-0.3cm} \includegraphics[width=0.35\textwidth]{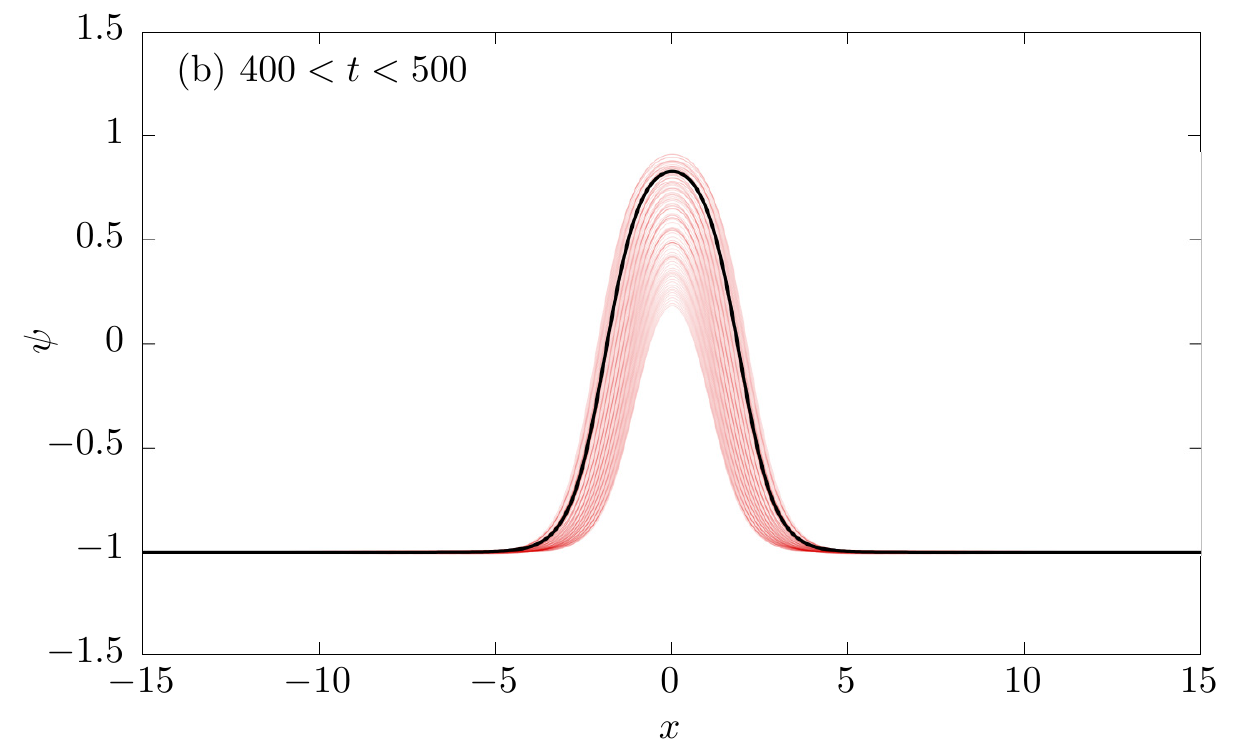}
\hspace*{-0.3cm} \includegraphics[width=0.35\textwidth]{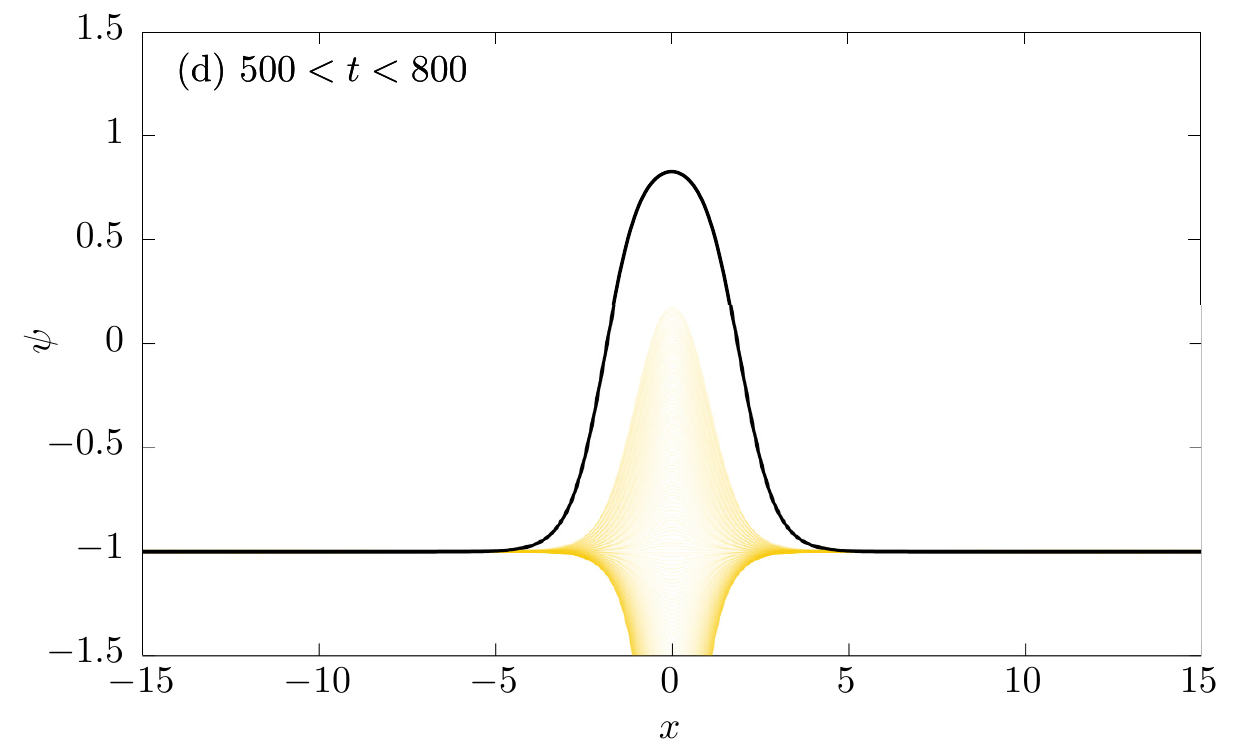}
\caption{Evolution of the field $\psi$ obtained in the scattering of the initial configuration (\ref{init-phi})-(\ref{init-psi}) with the purple mode $(n=6)$ initially excited. Here $A=0.05$ and solitons pass the SW. }
\label{purple-wall-1}
\vspace*{0.3cm}
\hspace*{-0.8cm}\includegraphics[width=0.35\textwidth]{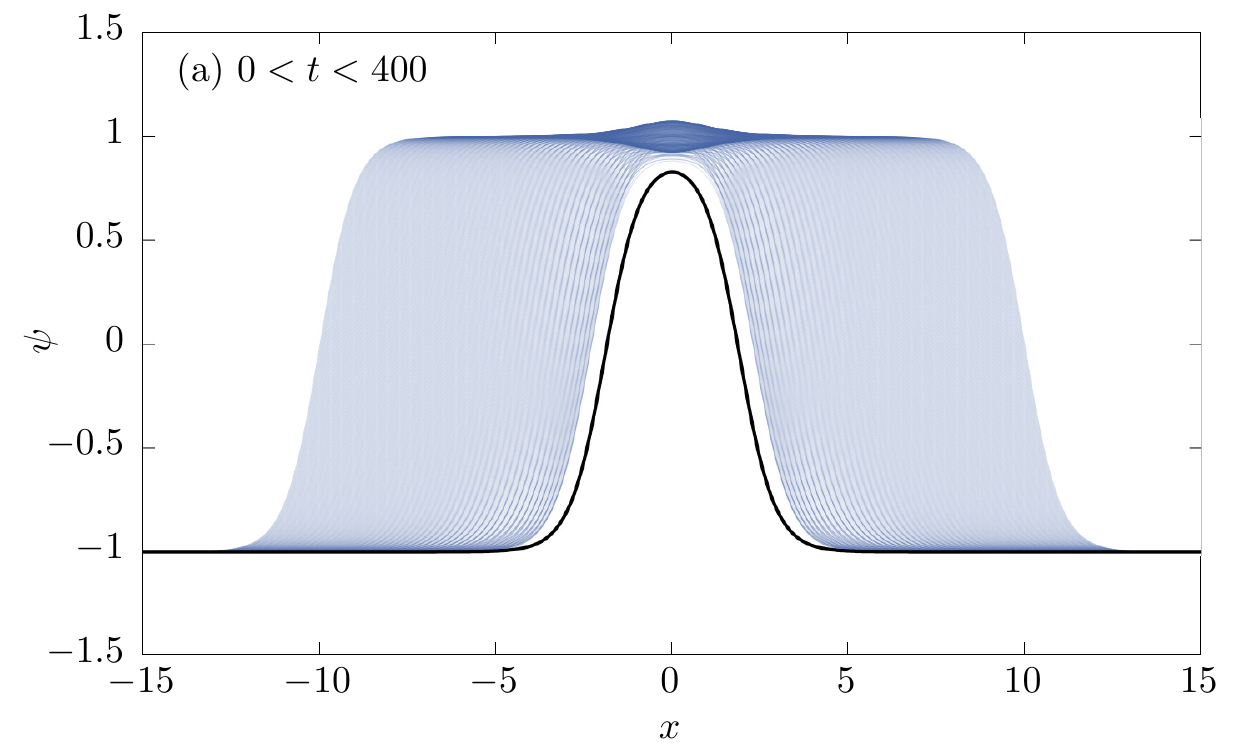}
\hspace*{-0.3cm} \includegraphics[width=0.35\textwidth]{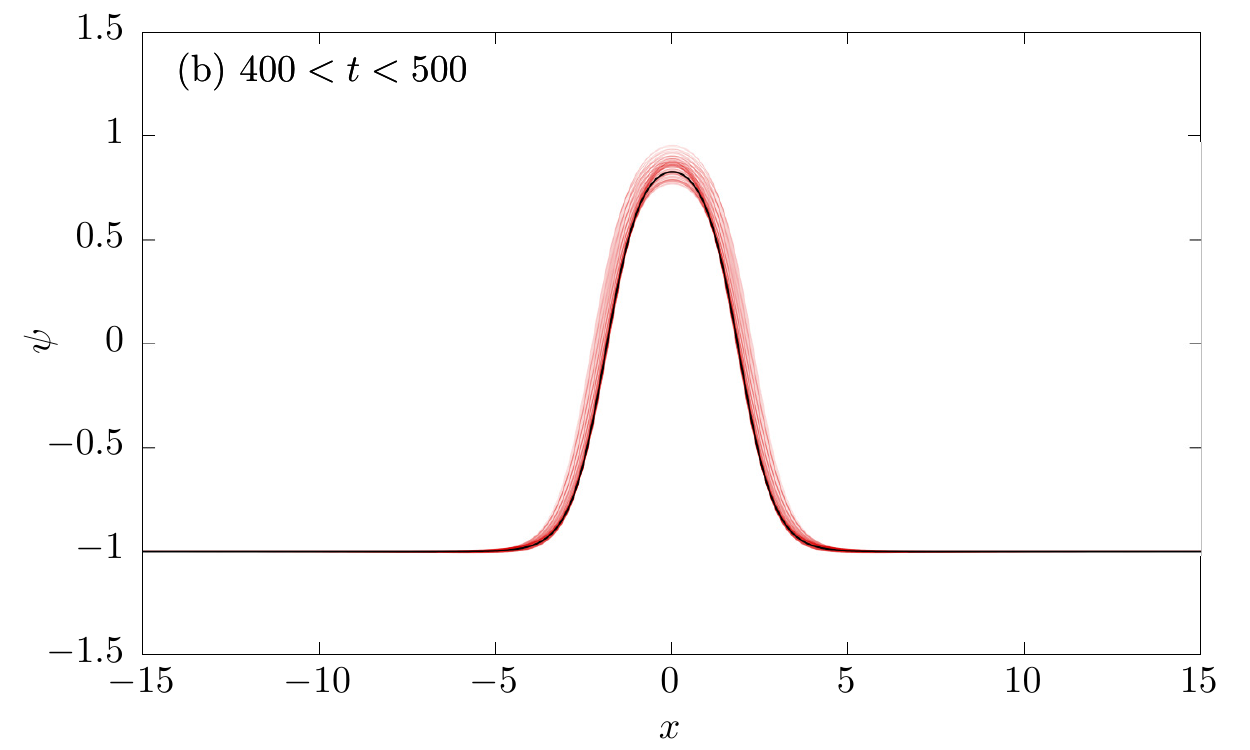}
\hspace*{-0.3cm} \includegraphics[width=0.35\textwidth]{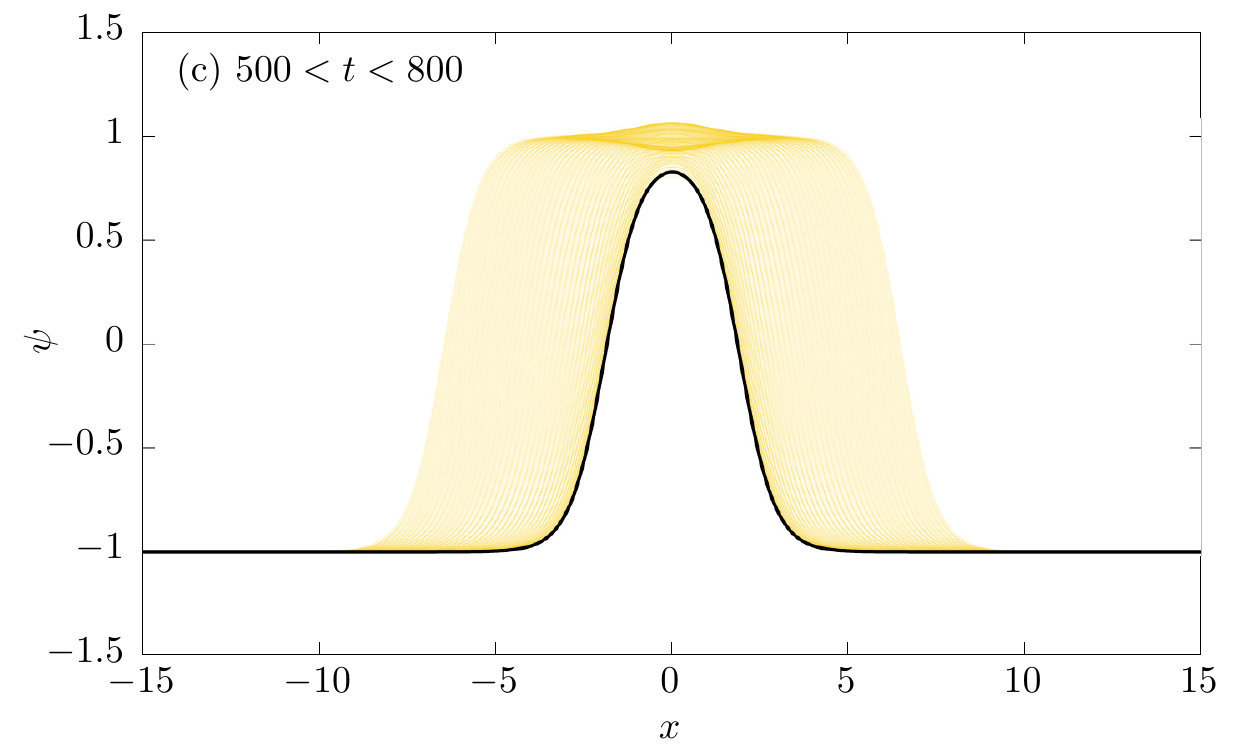}
\caption{Evolution of the field $\psi$ obtained in the scattering of the initial configuration (\ref{init-phi})-(\ref{init-psi}) with the purple mode  $(n=6)$ initially excited. Here $A=0.075$ and solitons form a long living stationary solution ($t \in [400,550]$) localized at the BPS solution with $b=10.63$. }
\label{purple-wall-2}
\vspace*{0.3cm}
 \hspace*{-0.8cm}\includegraphics[width=0.35\textwidth]{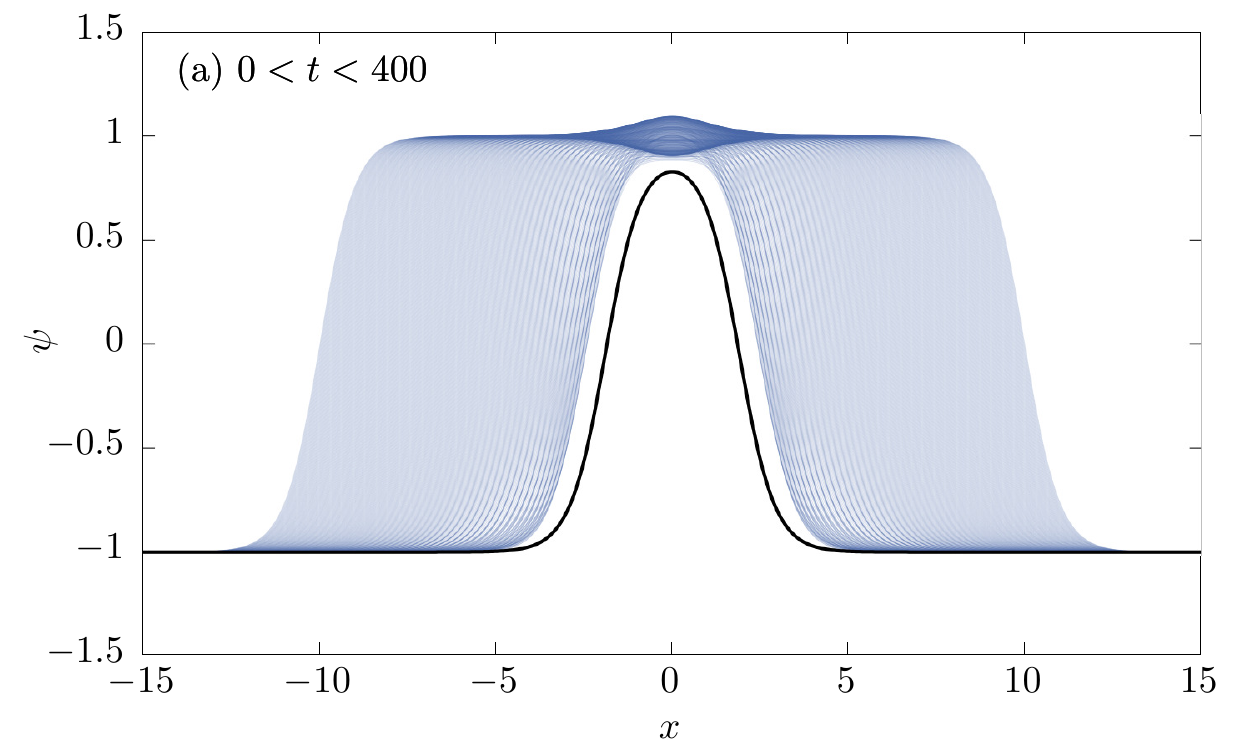}
\hspace*{-0.3cm} \includegraphics[width=0.35\textwidth]{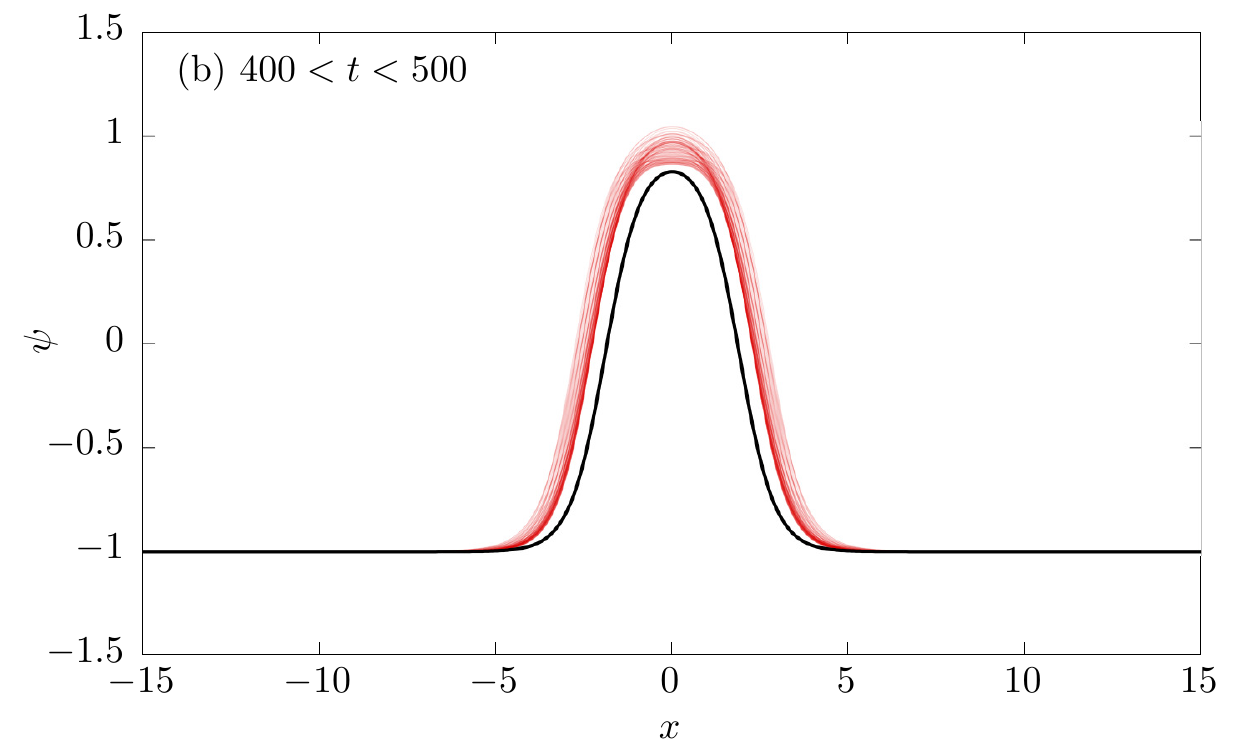}
\hspace*{-0.3cm} \includegraphics[width=0.35\textwidth]{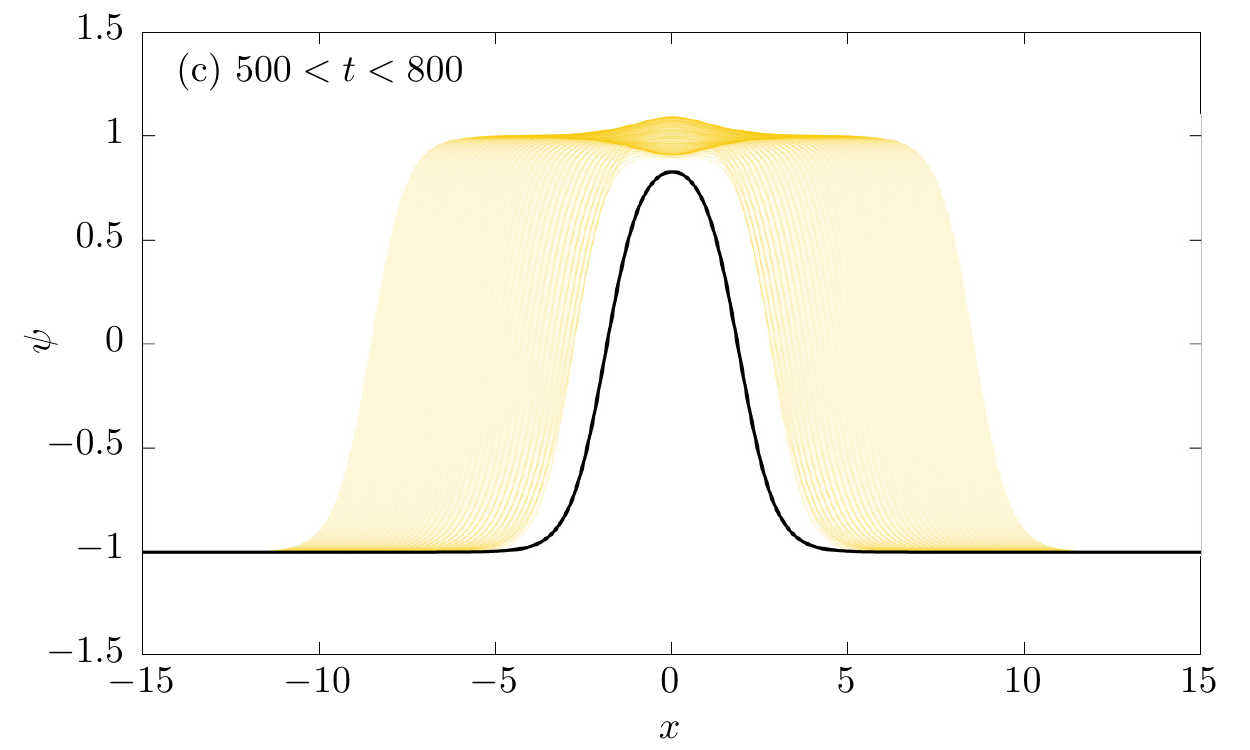}
\caption{Evolution of the field $\psi$ obtained in the scattering of the initial configuration (\ref{init-phi})-(\ref{init-psi}) with the purple mode  $(n=6)$ initially excited. Here $A=0.095$ and solitons are back scattered before the SW.}
\label{purple-wall-3}
\end{figure}

In Figs. \ref{pink-wall-1}-\ref{pink-wall-3}  we present snapshots of the corresponding profiles of the $\psi$ field evolved in the full (PDE) equations of motion for three representative values of the amplitude of the mode. The initial velocity is $v=0.05$ and $x_0=10$. The corresponding thin spectral wall should appear at the point where the mode crosses the mass threshold i.e., at $b=321$. This corresponds to a BPS solution (\ref{Q1}), which we have plotted as a black solid curve. 
In Fig. \ref{pink-wall-1} the mode amplitude is sufficiently small, $A=0.04$, that the incoming solitons of the $\psi$ field pass the wall (i.e., the BPS configuration with $b=321$). They also pass the SW of the other modes. This is not surprising as the SW is a selective phenomenon affecting only solutions with the corresponding mode excited. In Fig. \ref{pink-wall-2} the amplitude of the mode is equal to the critical value $A=0.078$. Now, we see very clearly that the incoming solitons approaching each other (panel (a) with time $t \in [0,300]$), form a stationary solution. Indeed, the field profile (and, of course, the positions of the solitons i.e., the zeros of the field $\psi$) freezes exactly at the relevant BPS solution (panel (b) with $t \in [400,700]$). Finally, the kink and antikink are back scattered (panel (c) for $t\in [700, 1200]$). In Fig. \ref{pink-wall-3} the amplitude of the mode is high enough, $A=0.085$, to provide a back scattering of the incoming $\psi$ field solitons before they reach the stationary solution. All these scenarios agree completely with the previous results obtained within the framework of the BPS impurity models \cite{spectral-wall}, \cite{SD-phi4}. 

Next, we excite the purple mode, $(n=6)$, being a mode initially located at the origin, where the anitkink in the $\phi$ field is located. So we take
\bea
\phi_{in}&=& -\tanh x,\label{init-phi-6} \\
\psi_{in}&=& \tanh (x+x_0-vt) -  \tanh(x-x_0+vt) -1 +\frac{A}{\cosh x}.   \label{init-psi-6}
\eea
 The situation completely repeats itself with the obvious modification of the position of the SW. Now, the SW is located at $b=10.63$, i.e., at the point where the purple mode crosses the mass threshold; see Figs. \ref{purple-wall-1}-\ref{purple-wall-3}. Note that the SW related to the pink mode is completely transparent in this case; exactly as the purple SW was invisible for the case of the excited pink mode. This demonstrates the very selective nature of the spectral walls. 
\section{Singularities of the vibrational moduli space}
As mentioned before, the dynamical vibrational moduli space collapses at the value of the coordinates where a normal mode enters the continuum spectrum, i.e., at the position of the spectral wall. Such a collapse is well understood since the relevant collective coordinate, that is, the amplitude of the corresponding normal mode, ceases to exist. From the point of view of the metric on the vibrational moduli space it leads to the appearance of singularities which are accessible after a finite amount of time. Hence, the vibrational moduli space is not metrically complete. 

Again, we would like to stress that the collapse of the dynamical vibrational moduli space is due to the correct, dynamical treatment of the normal modes, which fully takes into account their change as we change the moduli $a^i$. One may think that the appearance of singularities suggests that the frozen vibrational moduli space would be a better choice for the vibrational moduli space. Indeed, the resulting metric does not have any singularities. However, the frozen vibrational moduli space is blind to the fact that the normal modes pass the mass threshold and therefore cannot explain the existence of the spectral walls.

\begin{figure}
\includegraphics[width=1.0\textwidth]{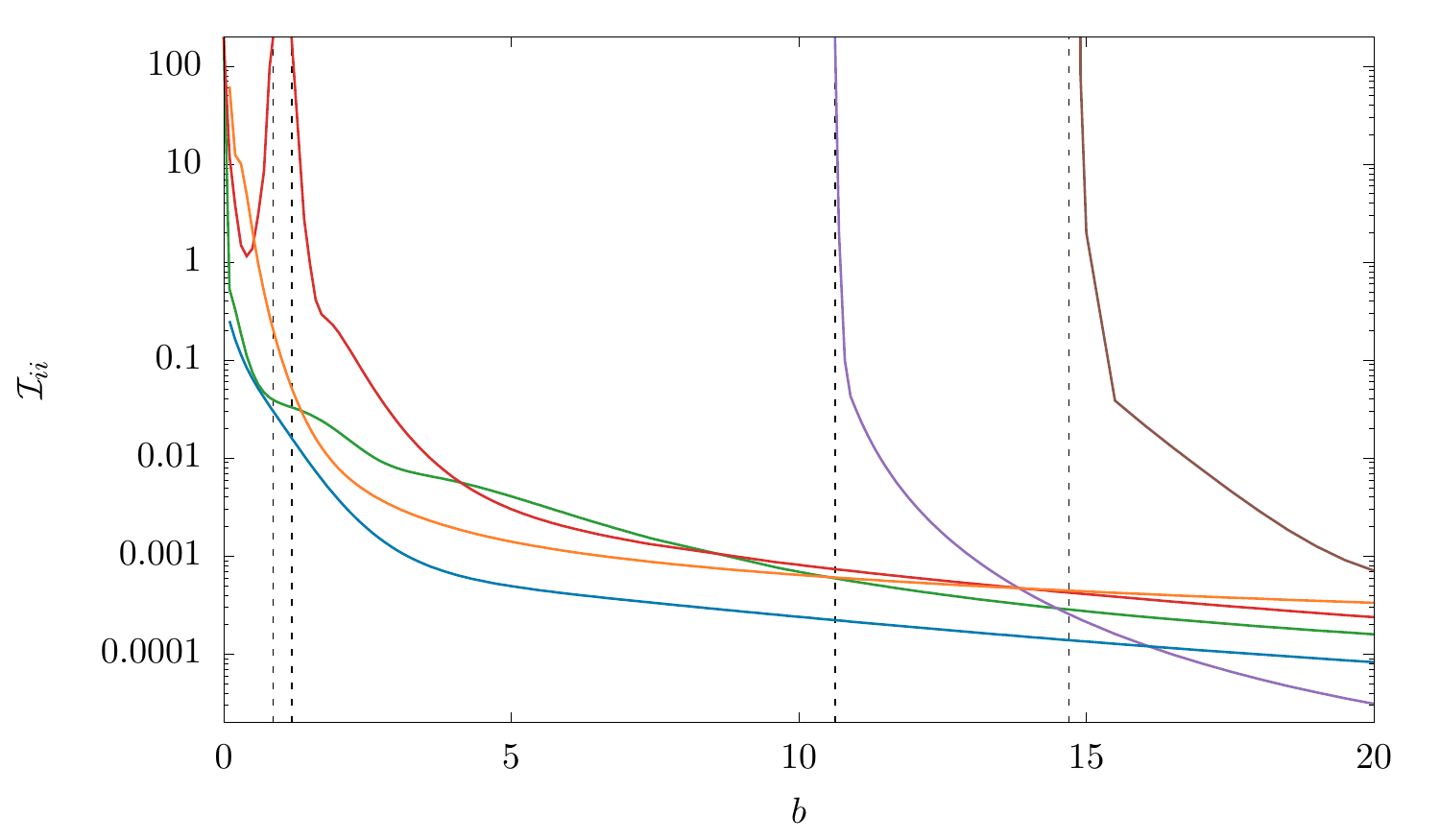}
\caption{The dependence of the couplings $\mathcal{I}_{ii}$ on the modulus $b$. Dashed vertical line denotes the positions of the spectral wall related to the $i$-th mode where the pertinent integral $\mathcal{I}_{ii}$ diverges. Here $i=7, 6, 5, 4$ which corresponds to the brown, purple, red and green modes respectively. Note that the red mode ($i=5$) leads to three singularities as it is expected from the previously found spectral structure, see Fig. \ref{spectral}.}
\label{singularities}
\end{figure}

As before, the metric on the dynamical vibrational moduli space is obtained by inserting the field configuration 
$\tilde{\Phi}(x,t)=\Phi_0(x;a^i)+\sum_{k=1}^M A^k\zeta_k(x;a^i)$
into the time dependent part of the Lagrangian with the assumption that the collective coordinates, $X^m=(a^i, A^k)$, depend on time. Thus,
\be
g_{mn}= \int_{-\infty}^{\infty} \frac{\partial \tilde\Phi}{\partial X^m} \frac{\partial \tilde\Phi}{\partial X^n} dx.
\ee
In the case considered here we take the BPS solutions with all the possible orthonormal bound modes: 
\bea
\tilde\phi(x,t)&=& \phi_0^-(x;a)+\sum_{k=1}^7 A^k \eta_k(x;a,b), \label{vib-1} \\
\tilde\psi (x,t)&=& \psi_0^-(x;a,b)+\sum_{i=k}^7 A^k \xi_k(x;a,b), \label{vib-2}
\eea
where $a,b, A^k$ are functions of time. This provides a dynamical vibrational moduli space with nine coordinates. For reasons of simplicity, we truncate it to an eight-dimensional subspace assuming that $a=\mbox{const.}=0$. Hence, $X^m\equiv (X^0,X^k)=(b,A^k)$, where $m=0...7$ and $k=1...7$. This means that the centre of mass of the scattering solitons does not move. This truncation, considered in all our numerical computations, is a consistent truncation based on the observation that the eigenmodes and their frequencies do not depend on $a$. The resulting metric components read
\bea
g_{00}&=& \left( g_{bb} + \mathcal{I}_{ij} A^i A^j + \mathcal{I}_i A^i \right), \\
g_{kl} &=& \delta_{kl}, \\
g_{0k} &=& \mathcal{K}_{kl} A_l,
\eea
where $g_{bb}$ is the part already obtained for the canonical moduli space (\ref{gbb}), while 
\be
\mathcal{I}_{ij} = \int_{-\infty}^{\infty} \left( \frac{d \eta_i}{d b} \frac{d \eta_j}{d b} + \frac{d \xi_i}{d b} \frac{d \xi_j}{d b} \right) dx,
\ee
 \be
 \mathcal{I}_{i} = \int_{-\infty}^{\infty}  \frac{d \psi_0}{d b} \frac{d \xi_i}{d b} dx,
 \ee 
 and
 \be
  \mathcal{K}_{kl} =  \int_{-\infty}^{\infty}  \left( \eta_i \frac{d \eta_j}{d b} +  \xi_i \frac{d \xi_j}{d b} \right) dx.
 \ee
 In Fig. \ref{singularities} we show that there are terms in the metric component $g_{00}$ which diverge at $b$ values at which  a normal mode enters the continuum. Specifically, the integrals $\mathcal{I}_{ii}$ diverge at $b=b^{(i)}_{sw}$, where $i$ is the number of the massive normal mode and $b^{(i)}_{sw}$ the value(s) of the modulus $b$ at which it crosses the mass threshold. In Fig. \ref{singularities} we demonstrate this for the brown ($i=7$), purple ($i=6$), red ($i=5$) and green ($i=4$) modes. We see clearly that the singularities of the metric coincide with the position of the spectral walls. The integral $\mathcal{I}_{55}$ (red mode) diverges three times as expected from the spectral structure behaviour (see Fig. \ref{spectral}).

We would like to remark that the computation of the coefficients of the effective model, and specifically the integrals $\mathcal{I}_{ii}$, was a very delicate numerical problem. Given the fact that the eigensolutions near the threshold were found with controllable but significant errors, the main issue was to calculate accurately the derivatives $\partial\xi/\partial b$ and $\partial\eta/\partial b$ which, from a numerical point of view, required a subtraction of two similar values. This typically leads to large relative errors of the result (so-called cancellation problem). 
We hope that we will be able to improve our method in the future using some other numerical techniques and employ a perturbative approach.

\vspace*{0.2cm}

Of course, the singularities of the dynamical vibrational moduli space lead to the breakdown of the effective model obtained by the insertion of the restricted field configurations (\ref{vib-1})-(\ref{vib-2}) into the original Lagrangian. Indeed, the effective model consists of the kinetic term, which takes into account the metric on the dynamical vibrational moduli space, supplemented by a potential (non-derivative) part $U$
\be
L[b,A^1,..,A^7]=\frac{1}{2} g_{00} \dot{b}^2 +\frac{1}{2}  \dot{A}_k^2 +g_{0k} \dot{b}\dot{A}^k- U(b, A^1..A^7) 
\ee
where
\bea
& & \hspace*{-1.8cm} U(b, A^1,..,A^7) = \\
& & \int_{-\infty}^\infty \left[ \frac{1}{2} \left(\frac{d\tilde{\phi}}{dx} \right)^2 +\frac{1}{2} \left(\frac{d\tilde\psi}{dx} \right)^2 + \frac{1}{2} (\tilde\phi^2-1)^2 + \frac{1}{2}  \tilde\phi^2(\tilde\psi^2-1)^2- \tilde\phi (1-\tilde\psi^2)\frac{d\tilde\psi}{dx}  \right] dx
\eea
is a rather complicated, but regular, function of $b$, which is a polynomial in the amplitudes $A^k$. Thus, whenever the metric $g_{mn}$ possesses a singularity, the effective theory breaks down. Here again, for simplicity we assumed $a=0$. 

The collapse of the dynamical vibrational moduli space description reflects the method of the construction of this space of restricted field configurations. Namely, only normal modes (both zero and massive ones) have been used. However, after reaching the mass threshold, the normal mode does not completely disappear but transmutes into a quasi-normal mode. This observation opens a way to cure the singularities of the vibrational moduli space. One should build a collective coordinate effective model (based on an improved dynamical vibrational moduli space) using vibrational modes. This cannot be a trivial replacement of a normal mode by the corresponding quasi-normal mode when it enters the continuum. Quasi-normal modes are {\it non-normalizable} solutions of the spectral problem. Another approach is required which would treat normal and quasi-normal modes in the same manner. Then, we could obtain the critical amplitude at which the stationary solution, living on the spectral wall, is formed. Furthermore, it may be necessary to go beyond the linear perturbation theory to fully explain the behaviour of the BPS solitons in the vicinity of the spectral walls. This issue definitely requires further studies.

\section{The BPS-impurity as frozen defect}
Finally, we would like to show that there is a limit in which model (\ref{model}) approaches the BPS impurity model \cite{SD-phi4}. To do this we assume $m=1$ while the energy scale of the field $\phi$ is controlled by a parameter  $M$
\be
\mathcal{L}[\phi,\psi]= \frac{M}{2} \left(\partial_\mu \phi \right)^2 +\frac{1}{2} \left(\partial_\mu \psi \right)^2 - \frac{M}{2} (\phi^2-1)^2 - \frac{1}{2}  \phi^2(\psi^2-1)^2+ \phi (1-\psi^2)\partial_x \psi. 
\ee
In the limit $M \to \infty$, the field $\phi$ decouples from $\psi$ and obeys the usual $\phi^4$ equation of motion. In the static limit we get the solutions
\be
\phi_0^\infty = \pm 1 \;\;\;  \mbox{or} \;\;\; \phi_0^\infty = \pm \tanh (x+a),
\ee
where the first solution represents the vacua (zero energy). The second solution is an infinitely heavy, {\it frozen} kink or antikink of the $\phi^4$ theory. Inserting it back into the Lagrangian and neglecting the infinitely heavy part proportional to $M$ we get
\be
\mathcal{L}[\psi]= \frac{1}{2} \left(\partial_\mu \psi \right)^2 - \frac{1}{2}  \sigma^2(\psi^2-1)^2+ \sigma (1-\psi^2)\partial_x \psi, 
\ee
where $\sigma = \pm \tanh x$ is an impurity coupled in the BPS manner \cite{solvable-imp}. As we have already said, spectral walls in such a BPS impurity model are very well established \cite{SD-phi4}. The fact that the BPS impurity models \cite{solvable-imp} can be realized as a frozen defect of a theory with a larger target space resembles the situation in the Tong-Wong model of vortices and impurities \cite{tong}, \cite{krusch}, \cite{gud}.

\section{Summary}
In the present work we have confirmed, for the first time, the existence of thin spectral walls in a generic scalar field theory in (1+1) dimensions. Contrary to previous findings, there is no need for any impurity, i.e., a background field. A thin spectral wall is a far distance obstacle in the soliton dynamics which emerges due to the transition of a massive normal mode into the continuous spectrum during a solitonic collision. This phenomenon has two crucial properties. 

First of all, for a critical value of the mode amplitude it gives rise to the formation of a long living stationary state, i.e., a system of solitons frozen at a certain distance. This distance (the position of the SW) does not depend on particularities of the initial configuration (e.g., initial velocity of the solitons) but is given by a point where the vibrational manifold breaks down as one of the coordinates ceases to exist. This also explains why we call such SWs  {\it thin} SWs. If the amplitude is smaller than the critical value, solitons can pass the wall, while for larger amplitudes they are scattered back. The reflection point occurs sooner (for larger intersoliton distance) as the mode amplitude grows. 

Secondly, a SW is a very selective object. A spectral wall affects solitonic dynamics only if the relevant mode is excited. Hence, a SW related with one mode is transparent for solutions with other modes excited. 

As in the BPS impurity framework, we are able to predict the position of SWs but not the critical amplitude. This is because of the fact that, exactly at the location of a SW, the effective model or, more precisely, the dynamical vibrational moduli space, breaks down. Probably, a description which takes into account also non-normalizable, quasi-normal modes could handle this issue. However, this is a complicated and difficult problem which we leave for future investigations. In any case, another main result of the present paper is the observation that the singularities of the dynamical vibrational moduli space manifest as physical phenomena, i.e., spectral walls. Hence, we have shown that for excited BPS processes (scattering of multi-soliton BPS solutions with excited normal modes), despite its singularities, the dynamical vibrational moduli space provides a better framework for understanding the dynamics than the usual frozen vibrational moduli space. Indeed, in the latter case, the points of the location of the spectral walls, $b_{sw}$, do not play any particular role. 

It should be underlined that, although we chose a specific model for the identification and analysis of the thin spectral wall, this phenomenon will exist in other theories in (1+1) dimension with at least two scalar fields. The first necessary condition is just the existence of BPS solitons, i.e., topological solitons solving the corresponding Bogomol'nyi equations and therefore saturating the corresponding topological bound on the energy. As they are BPS solutions, the solitons can be located at arbitrary distance from each other. The energy is obviously independent of the mutual distance, but the spectral structure does change. Thus, the second necessary condition is that a mode crosses the mass threshold. This condition is not obeyed by all BPS models but, on the other hand, it is not a restrictive condition. It is known from the analysis of the BPS impurity models that there exists an infinitely large class of impurities (that is, frozen defects of the second field) with this property. 

Hence, we expect that thin spectral walls will have a great impact on the scattering processes in already known BPS systems of two real scalars in (1+1) dimensions. Here we mention the Montonen-Sarker-Trullinger-Bishop model \cite{m,stb} and its generalizations \cite{izq1,izq3}, two-component models arising in a Wess-Zumino model \cite{shif,bazeia2}, the BPS models studied in \cite{wojtek1}-\cite{gabriele} as well as models with a dielectric type coupling \cite{bazeia1} or other multiple scalar field theories \cite{multi-1}-\cite{multi-4}. 

There are no fundamental obstacles to the existence of thin spectral walls in higher dimensional systems with BPS solitons. Here the most natural example can be the Abelian Higgs model at critical coupling. Again, the spectral walls may be the main factor determining the vortex dynamics beyond the lowest order geodesic flow on the moduli space. 

An interesting question is the existence of spectral walls in two-scalar models which are not BPS theories, i.e., completely generic field theories in (1+1) dimensions, which of course have an even broader range of applications, see e.g.,  \cite{pnevmatikos}, \cite{non-BPS-1a}-\cite{izq6}. Here, again, the recent results obtained in the framework of the BPS impurity models can give us a hint. Namely, it was shown that in the near BPS-impurity models, when there is a very weak static force between kinks and impurity (frozen kink), infinitely thin spectral walls transmute into {\it thick spectral walls} \cite{near-SD-phi4}. This suggests that a stationary state is formed {\it before} the mode responsible for the spectral wall in the BPS case crosses the mass threshold, and its position does depend on the details of the initial state (e.g., velocity of the incoming solitons). Therefore, the mechanism behind the formation of such a thick wall is different. In this case it arises due to the balance between the kink-impurity attraction and a mode-induced repulsion. We conjecture that the same mechanism will also be present in weakly non-BPS two-scalar-field models. 

Finally, one may also ask about the importance of spectral walls in quantum theory where the vibrational modes can play a significant role, see e.g. \cite{Hal-vib}-\cite{Ev3}.

It should be stressed that our research was motivated by recent developments in the BPS (self-dual) impurity models. This simplified set-up allowed us to explain the impact of normal modes on the soliton dynamics in specific BPS theories. We believe that the BPS impurity models, providing a simplified but mathematically well controlled framework, may lead to further insights into the complicated problem of the interaction of topological solitons. 

We hope that our work will allow for a better understanding of the dynamics of topological solitons and in particular of the role played by the internal degrees of freedom. Furthermore, as spectral walls are rather generic phenomena existing in various models, we feel that they should soon be observed in experiments. 

\section*{Acknowledgements}

C.A and A.W. acknowledge financial support from the Ministry of Education, 
Culture, and Sports, Spain (Grant No. FPA2017-83814-P), the Xunta de 
Galicia (Grant No. INCITE09.296.035PR and Conselleria de Educacion) and 
the Spanish Consolider Program Ingenio 2010 CPAN (Grant No. 
CSD2007-00042). 
Further, this work has received financial support from Xunta de Galicia 
(Centro singular de investigación de Galicia accreditation 2019-2022), by 
the European Union ERDF, and by the “María de Maeztu” Units of Excellence 
program MDM-2016-0692 and the Spanish Research State Agency. K.O., T.R., and A.W. were supported by the Polish National Science Centre Grant No. NCN 2019/35/B/ST2/00059.

\end{document}